\documentclass[prd,nofootinbib,showpacs,superscriptaddress, preprint]{revtex4-1}

\usepackage[T1]{fontenc}
\usepackage{amsmath,amssymb}
\usepackage{epsfig}
\usepackage{graphicx}
\usepackage[usenames,dvipsnames]{color}
\usepackage{slashed}
\usepackage[colorlinks,citecolor=blue]{hyperref}
\usepackage{color}
\usepackage{subfig}

\begin{document}
\title{A New Viable Region of Inert Higgs Doublet Dark Matter Model with Scotogenic Extension}

\author{Debasish Borah}
\email{dborah@iitg.ernet.in}
\affiliation{Department of Physics, Indian Institute of Technology Guwahati, Assam 781039, India}
\author{Aritra Gupta}
\email{aritra@hri.res.in}
\affiliation{Harish-Chandra Research Institute, Chhatnag Road, Jhunsi, Allahabad 211019, India}
\affiliation{Homi Bhabha National Institute, Training School Complex,
Anushaktinagar, Mumbai - 400094, India}

\begin{abstract}
We explore the intermediate dark matter mass regime of the inert Higgs doublet model (IHDM), approximately between 400-550 GeV which is allowed by latest constraints from direct and indirect detection experiments, but the thermal relic abundance remains suppressed. We extend the model by three copies of right handed neutrinos, odd under the in built $Z_2$ symmetry of the model. This discrete $Z_2$ symmetry of the model allows these right handed neutrinos to couple to the usual lepton doublets through the inert Higgs doublet allowing the possibility of radiative neutrino mass in the scotogenic fashion. Apart from generating non-zero neutrino mass, such an extension can also revive the intermediate dark matter mass regime. The late decay of the lightest right handed neutrino to dark matter makes it possible for the usual thermally under-abundant dark matter in this intermediate mass regime to satisfy the correct relic abundance limit. The revival of this wide intermediate mass range can not only have relevance for direct and indirect search experiments but also for neutrino experiments as the long lifetime of the lightest right handed neutrino also results in almost vanishing lightest neutrino mass.
\end{abstract}
\maketitle

\section{Introduction}
The presence of non-baryonic form of matter, or the so called Dark Matter (DM) in large amount in the present Universe has been known to us for many decades. However, the nature of the dark matter particle is still a mystery. In spite of convincing and irrefutable experimental evidences suggesting the presence of DM in the Universe both from astrophysics and cosmological observations, all the experiments hunting for particle dark matter have given null results so far. Among these evidences, the galaxy cluster observations by Fritz Zwicky \cite{Zwicky:1933gu} back in 1933, observations of galaxy rotation curves in 1970's \cite{Rubin:1970zza}, the more recent observation of the bullet cluster \cite{Clowe:2006eq} and the latest cosmology data provided by the Planck satellite \cite{Ade:2015xua} are the most notable ones. The latest data from the Planck satellite indicate that approximately $27\%$ of the present Universe is composed of such mysterious non-baryonic dark matter, which is approximately five times more than the ordinary luminous or baryonic matter. The present abundance of DM is often quoted in terms of the density parameter $\Omega$ as \cite{Ade:2015xua}
\begin{equation}
\Omega_{\text{DM}} h^2 =\frac{\rho_{\text{DM}}}{\rho_{\text{c}}}h^2= 0.1186 \pm 0.0020
\label{dm_relic}
\end{equation}
where $h = H_0/100$ is a parameter of order unity and $\rho_{\text{c}}=\frac{3H^2_0}{8\pi G}$ is the critical density while $H_0$ being the present value of the Hubble parameter and $G$ being the universal constant of gravity. Although none of these experiments tell anything about the particle nature of DM, they have motivated the particle physics community to study DM within different particle physics models. The list of criteria which a particle dark matter candidate should fulfil \cite{Taoso:2007qk} rules out all the standard model (SM) particles from being dark matter candidates. This has led to several beyond standard model (BSM) proposals in the last few decades, the most popular of which being the so called weakly interacting massive particle (WIMP) paradigm. In this framework, a dark matter candidate typically with electroweak scale mass and interaction rate similar to electroweak interactions can give rise to the correct dark matter relic abundance, a remarkable coincidence often referred to as the \textit{WIMP Miracle}.

Since the WIMP interactions are of the order of electroweak interactions, they can be thermally produced in the early Universe. The relic abundance of such DM candidate can be generated as the interactions freezes out with the subsequent expansion as well as cooling of the Universe. The interesting part of this WIMP paradigm is that it also predicts observable DM nucleon scattering cross section through the same interactions that were operational at the time of freeze-out. However, no such DM nucleon scattering has been observed at different direct detection experiments. The most recent dark matter direct detection experiments like LUX, PandaX-II and XENON1T have also reported their null results \cite{Akerib:2016vxi, Tan:2016zwf, Aprile:2017iyp}. The absence of dark matter signals from the direct detection experiments have progressively lowered the exclusion curve in its mass-cross section plane. With such high precision measurements, the WIMP-nucleon cross section will soon overlap with the neutrino-nucleon cross section. Similar null results have been also reported by other direct search experiments like the large hadron collider (LHC) giving upper limits on DM interactions with the SM particles. A recent summary of collider searches for DM can be found in \cite{Kahlhoefer:2017dnp}. Although such null results could indicate a very constrained region of WIMP parameter space, they have also motivated the particle physics community to look for beyond the thermal WIMP paradigm. A new type of scenario that is well studied (outside the common thermal WIMP paradigm) belongs to the class of non-thermal dark matter candidates. These particles are often called as ``Feebly Interacting Massive Particles" or FIMPs \cite{Hall:2009bx}, due to extremely weak interaction strengths. It's due this weak coupling strengths, these particles can never become a part of the thermal plasma, and hence always remain out of equilibrium. Here, relic density is achieved by the so called ``freeze-in" mechanism in contrast to the usual freeze-out. In the former case, unlike freeze-out, the relic density depends on the details of its production history. The dark matter abundance in case of freeze-in rises from very small initial values (due to very low interaction strengths) to a final constant (relic) value. Well studied candidates in the category of non-thermal dark matters are axinos \cite{Covi:2001nw,Choi:2013lwa}, WIMPzillas \cite{Kolb:1998ki}, gravitinos \cite{Kallosh:1999jj,Giudice:1999am} and others. Production in these non-thermal scenarios are usually from the decays of heavier mother particles (e.g. from Inflaton and Moduli decays \cite{Endo:2006ix,Shaposhnikov:2006xi}).

Instead of completely giving up on the WIMP framework due to the null results at direct search experiments, here we consider a scenario where both thermal and non-thermal contributions to DM relic abundance can be important. To be more specific, DM can have sizeable interactions to be thermally produced in the early Universe but also needs a non-thermal source to satisfy the correct relic abundance. For concrete realisation of this scenario, we consider a minimal dark matter model which has been very widely studied in the literature. This model is a minimal extension of the SM by a Higgs field which transforms as a doublet under the $SU(2)_L$ gauge symmetry of the SM and has a hypercharge $Y=1$. An in-built discrete $Z_2$ symmetry under which this additional Higgs doublet has an odd charge while all SM fields have even charges guarantees the stability of the lightest component of the Higgs doublet. If the lightest component of this odd or inert Higgs doublet is electromagnetically neutral, it can be a good DM candidate. After being introduced by the authors of \cite{Deshpande:1977rw}, this inert Higgs doublet model (IHDM) has been extensively studied in the literature \cite{Ma:2006km,Barbieri:2006dq,Cirelli:2005uq, LopezHonorez:2006gr,Honorez:2010re, LopezHonorez:2010tb,Arhrib:2013ela, Dasgupta:2014hha}. Apart from predicting a stable WIMP dark matter candidate that can give rise to the correct relic abundance, the IHDM has several interesting implications that have been studied in the existing literature. For example, the promising indirect detection prospects of DM in this model have been studied in several works some of which can be found in \cite{Eiteneuer:2017hoh, Garcia-Cely:2015khw, Queiroz:2015utg, Garcia-Cely:2013zga}. Similarly, the direct search prospects at colliders like the LHC have been discussed in many works including \cite{Belyaev:2016lok, Hashemi:2016wup, Poulose:2016lvz, Datta:2016nfz, Belanger:2015kga, Ilnicka:2015jba, Goudelis:2013uca, Gustafsson:2012aj}. The possibility of a strong electroweak phase transition in this model have also been studied by several groups including \cite{Chowdhury:2011ga, Borah:2012pu, Gil:2012ya, Blinov:2015vma}. Such a phase transition, which is not possible in the SM, can have very interesting implications for producing the matter-antimatter asymmetry of the Universe. Since the SM gives rise to a metastable vacuum, the prospects of obtaining vacuum stability within IHDM was also studied in \cite{Chakrabarty:2015yia, Khan:2015ipa}.

The earlier works on IHDM have shown that there are primarily two regions of DM mass where the observed relic abundance can be produced: one below the $W$ boson mass threshold $(M_{\text{DM}} < M_W)$ and the other around 550 GeV or above. In the low mass region, as we discuss in details below, the tree level DM-SM interaction through the SM Higgs $(h)$ portal is interesting as it can simultaneously control the relic abundance as well as the DM-nucleon scattering cross section. In fact, this low mass region suffers from the strongest constraints from the direct detection experiments. For example, the latest data from the LUX experiment rules out DM-nucleon spin independent cross section above around $2.2 \times 10^{-46} \; \text{cm}^2$ for DM mass of around 50 GeV \cite{Akerib:2016vxi}. On the other hand, the recently released results from the XENON1T experiment rules out spin independent WIMP-nucleon interaction cross section above $7.7\times 10^{-47}\; \text{cm}^2$ for DM mass of 35 GeV \cite{Aprile:2017iyp}. These strong bounds reduce the allowed DM masses in the low mass regime to a very narrow region near the resonance $M_{\text{DM}} \approx m_h/2$. Although the direct detection limits can be somewhat relaxed in the high mass regime $(M_{\text{DM}}  \lessapprox 550 \; \text{GeV})$, the direct production of DM at colliders will be suppressed compared to the low mass regime. Here we consider the possibility of reviving the intermediate dark matter mass range $(M_W < M_{\text{DM}} \lessapprox 550 \; \text{GeV})$ by suitable extension of the model. We refer to this intermediate mass range of IHDM as \textit{IHDM Desert} where the observed relic abundance of DM can not be generated. Revival of this IHDM desert will be interesting for collider studies, in most of which this region has perhaps been overlooked due to unsuccessful production of DM relic abundance.

Our motivation to extend the IHDM is two-fold: firstly to revive the IHDM desert and to generate light neutrino masses which remains unaddressed in the pure IHDM. We also note that the IHDM desert can not be revived by including additional DM interactions contributing to the thermal relic abundance. This is because, in the IHDM desert the DM annihilation cross section is already more than required to produce correct abundance through the usual freeze-out mechanism, as we discuss in details below. Due to large annihilation rates, the DM is under-abundant in this range and hence we include a possible way of generating the correct abundance due to the late decay of a heavy particle into DM. Interestingly, such a possibility is naturally present in the scotogenic model \cite{Ma:2006km} which is an extension of the IHDM by three copies of $Z_2$ odd right handed neutrinos singlet under the SM gauge symmetry. The SM neutrinos which remain massless at tree level can acquire a non-zero mass at one loop with the DM and singlet neutral fermions going inside the loop. We then consider the possibility of the lightest right handed neutrino decaying into DM after the thermal freeze-out of DM and bring back the under-abundant DM relic into the observed range. It should be noted that, our scenario is different from the usual freeze-in scenarios mentioned above. In the freeze-in scenarios, the DM is very weakly interacting with the rest of the thermal bath and hence it never comes in thermal equilibrium with the latter. However, in the present scenario studied in this work, the DM was in thermal equilibrium with the rest of the cosmological plasma but its final relic abundance is small due to large annihilation. It is worth mentioning that the authors of
\cite{Molinaro:2014lfa}, in the later part of their work, pointed out this possibility of a right handed singlet neutrino decay to inert scalar doublet WIMP dark matter at late epochs, within the framework of scotogenic model. Here we study this scenario in details by solving the coupled Boltzmann equations numerically and scan the whole parameter space that can give rise to correct dark matter abundance in the intermediate mass regime of IHDM. Another interesting possibility to revive the IHDM desert is to consider a two-component DM scenario so that another DM component can fill the deficit created by the under-abundant inert Higgs doublet dark matter having mass in the IHDM desert \cite{Alves:2016bib}. We also test the viability of this IHDM desert in view of the latest constraints from direct and indirect detection experiments of DM. While DM in this mass region can easily evade direct detection constraints from the XENON1T experiment by suitable tuning of DM-Higgs coupling, the indirect detection constraints from the Fermi Large Area Telescope (LAT) \cite{Ahnen:2016qkx} are however, more severe as they can rule out a portion of IHDM desert $(M_W < M_{\text{DM}} \lessapprox 400 \; \text{GeV})$. In view of these latest constraints, we define the IHDM desert as $(400 \; \text{GeV} < M_{\text{DM}} \lessapprox 550 \; \text{GeV})$. This region is allowed from experimental constraints and IHDM alone can not produce the correct relic abundance. Therefore, we particularly focus on this mass range in the scotogenic model and study the possibility of generating correct relic abundance from a non-thermal component. This region is not only very sensitive to the indirect detection experiments, but will also increase the collider prospects compared to the high mass regime.

The requirement of the long-livedness of the lightest right handed neutrino forces the corresponding Yukawa coupling with the DM and neutrino to be very small. This results in an almost vanishing light neutrino mass that can be probed at experiments sensitive to the absolute neutrino mass scale, like the neutrinoless double beta decay $(0\nu \beta \beta)$ experiments. We show that all the IHDM desert can be revived in this fashion and DM relic abundance can be successfully produced along with satisfying neutrino oscillation data. The opening up of this new DM mass range should invite a detailed study of collider prospects of this model specifically focusing on the IHDM desert. 

This work is organised as follows. In section \ref{sec1}, we briefly discuss the inert Higgs doublet model followed by a study of dark matter phenomenology of it in section \ref{sec2}. In section \ref{sec3} we discuss the scotogenic version of this model with radiative neutrino mass followed by the corresponding dark matter phenomenology in section \ref{sec4}. We finally conclude in section \ref{sec5}.

\section{The Inert Higgs Doublet Model}
\label{sec1}
As mentioned in the previous section, the inert Higgs doublet model (IDM) \cite{Barbieri:2006dq,Cirelli:2005uq,LopezHonorez:2006gr} is one of the simplest extensions of the SM in order to accommodate a DM candidate. The symmetry of the SM is extended by an additional global discrete $Z_2$ symmetry under which a newly incorporated scalar doublet $\Phi_2$ transforms as $\Phi_2 \rightarrow -\Phi_2$. The usual SM fields have even $Z_2$ charges and hence transform as $\Phi_{\text{SM}} \rightarrow \Phi_{\text{SM}}$. This $Z_2$ symmetry, though ad-hoc in this minimal setup, could be realised naturally as a subgroup of a continuous gauge symmetry like $U(1)_{B-L}$ with non-minimal field content \cite{Dasgupta:2014hha}. The $Z_2$ symmetry and the corresponding charges of the fields prevent the SM fermion couplings with the additional scalar $\Phi_2$ at renormalisable level making it inert, as the name of the model suggests. This also prevents linear and trilinear couplings of $\Phi_2$ with the SM Higgs. Therefore, if the bare mass squared term of $\Phi_2$ is chosen positive definite, its neutral components do not acquire any vacuum expectation value (vev) even after electroweak symmetry breaking (EWSB). This ensures the stability of the lightest component of $\Phi_2$, irrespective of its mass, on cosmological scale. If this lightest component is electromagnetically neutral, then this can be a good DM candidate, if other relevant constraints are satisfied. The scalar potential of the model involving the SM Higgs doublet $\Phi_1$ and the inert doublet $\Phi_2$ can be written as
\begin{equation}
\begin{aligned}
V(\Phi_1,\Phi_2) &=  \mu_1^2|\Phi_1|^2 +\mu_2^2|\Phi_2|^2+\frac{\lambda_1}{2}|\Phi_1|^4+\frac{\lambda_2}{2}|\Phi_2|^4+\lambda_3|\Phi_1|^2|\Phi_2|^2 \nonumber \\
& +\lambda_4|\Phi_1^\dag \Phi_2|^2 + \{\frac{\lambda_5}{2}(\Phi_1^\dag \Phi_2)^2 + \text{h.c.}\}
\end{aligned}
\label {c}
\end{equation}
To ensure that none of the neutral components of the inert Higgs doublet acquire a non-zero vev $\mu_2^2 >0$ is assumed. This also prevents the $Z_2$ symmetry from being spontaneously broken. The electroweak symmetry breaking occurs due to the non-zero vev acquired by the neutral component of $\Phi_1$. After the EWSB these two scalar doublets can be written in the following form in the unitary gauge.
\begin{equation}
\Phi_1=\begin{pmatrix} 0 \\  \frac{ v +h }{\sqrt 2} \end{pmatrix} , \Phi_2=\begin{pmatrix} H^\pm\\  \frac{H+iA}{\sqrt 2} \end{pmatrix}
\end{equation}
The masses of the physical scalars at tree level can be written as
\begin{eqnarray}
m_h^2 &=& \lambda_1 v^2 ,\nonumber\\
m_{H^\pm}^2 &=& \mu_2^2 + \frac{1}{2}\lambda_3 v^2 , \nonumber\\
m_{H}^2 &=& \mu_2^2 + \frac{1}{2}(\lambda_3+\lambda_4+\lambda_5)v^2=m^2_{H^\pm}+
\frac{1}{2}\left(\lambda_4+\lambda_5\right)v^2  , \nonumber\\
m_{A}^2 &=& \mu_2^2 + \frac{1}{2}(\lambda_3+\lambda_4-\lambda_5)v^2=m^2_{H^\pm}+
\frac{1}{2}\left(\lambda_4-\lambda_5\right)v^2.
\label{mass_relation}
\end{eqnarray}
Here $m_h$ is the SM like Higgs boson mass, $m_H, m_A$ are the masses of the CP even and CP odd scalars from the inert doublet. $m_{H^\pm}$ is the mass of the charged scalar. Without any loss of generality, we consider $\lambda_5 <0, \lambda_4+\lambda_5 <0$ so that the CP even scalar is the lightest $Z_2$ odd particle and hence a stable dark matter candidate.

The new scalar fields discussed above can be constrained from the LEP I precision measurement of the $Z$ boson decay width. In order to forbid the decay channel $Z \rightarrow H A$, one arrives at the constraint $m_H + m_A > m_Z$. In addition to this, the LEP II constraints roughly rule out the triangular
region \cite{Lundstrom:2008ai}
\[
	m_{H} < 80\ {\rm\ GeV},\quad m_{A} < 100{\rm\ GeV},\quad
	m_{A} - m_{H} > 8{\rm\ GeV}
\]
The LEP collider experiment data restrict the charged scalar mass to $m_{H^\pm} > 70-90$ GeV \cite{Pierce:2007ut}. The Run 1 ATLAS dilepton limit is discussed in the context of IHDM in Ref.\cite{Belanger:2015kga} taking into consideration of specific masses of charged Higgs. Another important restriction on $m_{H^\pm}$ comes from the electroweak precision data (EWPD). Since the contribution of the additional doublet $\phi_2$ to electroweak S parameter is always small \cite{Barbieri:2006dq}, we only consider the contribution to the electroweak T parameter here. The relevant contribution is given by \cite{Barbieri:2006dq}
\begin{equation}
\Delta T = \frac{1}{16 \pi^2 \alpha v^2} [F(m_{H^\pm}, m_{A})+F(m_{H^\pm}, m_{H}) -F(m_{A}, m_{H})]
\end{equation}
where 
\begin{equation}
F(m_1, m_2) = \frac{m^2_1+m^2_2}{2}-\frac{m^2_1m^2_2}{m^2_1-m^2_2} \text{ln} \frac{m^2_1}{m^2_2}
\end{equation}
The EWPD constraint on $\Delta T$ is given as \cite{LopezHonorez:2010tb}
\begin{equation}
-0.1 < \Delta T + T_h < 0.2
\end{equation}
where $T_h \approx -\frac{3}{8 \pi \cos^2{\theta_W}} \text{ln} \frac{m_h}{m_Z}$ is the SM Higgs contribution to the T parameter \cite{Peskin:1991sw}. 

\section{Dark Matter in IHDM}
\label{sec2}
The relic abundance of a dark matter particle $\rm DM$, which was in thermal equilibrium at some earlier epoch can be calculated by solving the Boltzmann equation
\begin{equation}
\frac{dn_{\rm DM}}{dt}+3Hn_{\rm DM} = -\langle \sigma v \rangle (n^2_{\rm DM} -(n^{\rm eq}_{\rm DM})^2)
\end{equation}
where $n_{\rm DM}$ is the number density of the dark matter particle $\rm DM$ and $n^{\rm eq}_{\rm DM}$ is the number density when $\rm DM$ was in thermal equilibrium. $H$ is the Hubble expansion rate of the Universe and $ \langle \sigma v \rangle $ is the thermally averaged annihilation cross section of the dark matter particle $\rm DM$. In terms of partial wave expansion $ \langle \sigma v \rangle = a +b v^2$. Numerical solution of the Boltzmann equation above gives \cite{Kolb:1990vq,Scherrer:1985zt}
\begin{equation}
\Omega_{\rm DM} h^2 \approx \frac{1.04 \times 10^9 x_F}{M_{\text{Pl}} \sqrt{g_*} (a+3b/x_F)}
\end{equation}
where $x_F = M_{\rm DM}/T_F$, $T_F$ is the freeze-out temperature, $M_{\rm DM}$ is the mass of dark matter, $g_*$ is the number of relativistic degrees of freedom at the time of freeze-out and and $M_{\text{Pl}} \approx 2.4\times 10^{18}$ GeV is the Planck mass. Dark matter particles with electroweak scale mass and couplings freeze out at temperatures approximately in the range $x_F \approx 20-30$. More generally, $x_F$ can be calculated from the relation 
\begin{equation}
x_F = \ln \frac{0.038gM_{\text{Pl}}M_{\rm DM}<\sigma v>}{g_*^{1/2}x_F^{1/2}}
\label{xf}
\end{equation}
which can be derived from the equality condition of DM interaction rate $\Gamma = n_{\rm DM} \langle \sigma v \rangle$ with the rate of expansion of the Universe $H \approx g^{1/2}_*\frac{T^2}{M_{Pl}}$. There also exists a simpler analytical formula for the approximate DM relic abundance \cite{Jungman:1995df}
\begin{equation}
\Omega_{\rm DM} h^2 \approx \frac{3 \times 10^{-27} cm^3 s^{-1}}{\langle \sigma v \rangle}
\label{eq:relic}
\end{equation}
The thermal averaged annihilation cross section $\langle \sigma v \rangle$ is given by \cite{Gondolo:1990dk}
\begin{equation}
\langle \sigma v \rangle = \frac{1}{8m^4T K^2_2(m/T)} \int^{\infty}_{4m^2}\sigma (s-4m^2)\surd{s}K_1(\surd{s}/T) ds
\end{equation}
where $K_i$'s are modified Bessel functions of order $i$, $m$ is the mass of Dark Matter particle and $T$ is the temperature.

Here we consider one of the neutral component of the scalar doublet $\Phi_2$ namely, $H$ (not to be confused with the Hubble parameter) as the dark matter candidate ($H \equiv \rm DM$) which is similar to the inert doublet model of dark matter discussed extensively in the literature \cite{Ma:2006km,Barbieri:2006dq,Cirelli:2005uq, LopezHonorez:2006gr,Honorez:2010re, LopezHonorez:2010tb,Arhrib:2013ela, Dasgupta:2014hha}. In the low mass regime $m_H \equiv M_{\text{DM}} \leq M_W$, dark matter annihilation into the SM fermions through s-channel Higgs mediation dominates over other channels. As pointed out by \cite{Honorez:2010re}, the dark matter annihilations $H H \rightarrow W W^* \rightarrow W f \bar{f^{\prime}}$ can also play a role in the $M_{\text{DM}} \leq M_W$ region. Also, depending on the mass differences $m_{H^\pm}-m_H\,(\equiv \Delta M_{H^\pm}), m_A-m_H\,( \equiv \Delta M_A)$, the coannihilations of $H, H^\pm$ and $H, A$ can also play a role in generating the relic abundance of dark matter. Specially when the heavier components of the inert scalar doublet have masses close to the DM mass, they can be thermally accessible at the epoch of DM freeze-out. Therefore, the annihilation cross section of dark matter in such a case gets additional contributions from coannihilations between dark matter and heavier components of the scalar doublet $\Phi_2$. This type of coannihilation effects on dark matter relic abundance were studied by several authors in \cite{Griest:1990kh, Edsjo:1997bg, Bell:2013wua}. Here we summarise the analysis of \cite{Griest:1990kh} for the calculation of the effective annihilation cross section in such a case. The effective cross section can given as 
\begin{align}
\sigma_{eff} &= \sum_{i,j}^{N}\langle \sigma_{ij} v\rangle r_ir_j \nonumber \\
&= \sum_{i,j}^{N}\langle \sigma_{ij}v\rangle \frac{g_ig_j}{g^2_{eff}}(1+\Delta_i)^{3/2}(1+\Delta_j)^{3/2}e^{\big(-x_F(\Delta_i + \Delta_j)\big)} \nonumber \\
\end{align}
where $x_F = \frac{m_{DM}}{T_F}$ and $\Delta_i = \frac{m_i-M_{\text{DM}}}{M_{\text{DM}}}$  and 
\begin{align}
g_{eff} &= \sum_{i=1}^{N}g_i(1+\Delta_i)^{3/2}e^{-x_F\Delta_i}
\end{align}
The masses of the heavier components of the inert Higgs doublet are denoted by $m_{i}$. The thermally averaged cross section can be written as
\begin{align}
\langle \sigma_{ij} v \rangle &= \frac{x_F}{8m^2_im^2_jM_{\text{DM}}K_2((m_i/M_{\text{DM}})x_F)K_2((m_j/M_{\text{DM}})x_F)} \times \nonumber \\
& \int^{\infty}_{(m_i+m_j)^2}ds \sigma_{ij}(s-2(m_i^2+m_j^2)) \sqrt{s}K_1(\sqrt{s}x_F/M_{\text{DM}}) \nonumber \\
\label{eq:thcs}
\end{align}
\begin{figure}[h!]
\centering
$
\begin{array}{cc}
\includegraphics[scale=0.4]{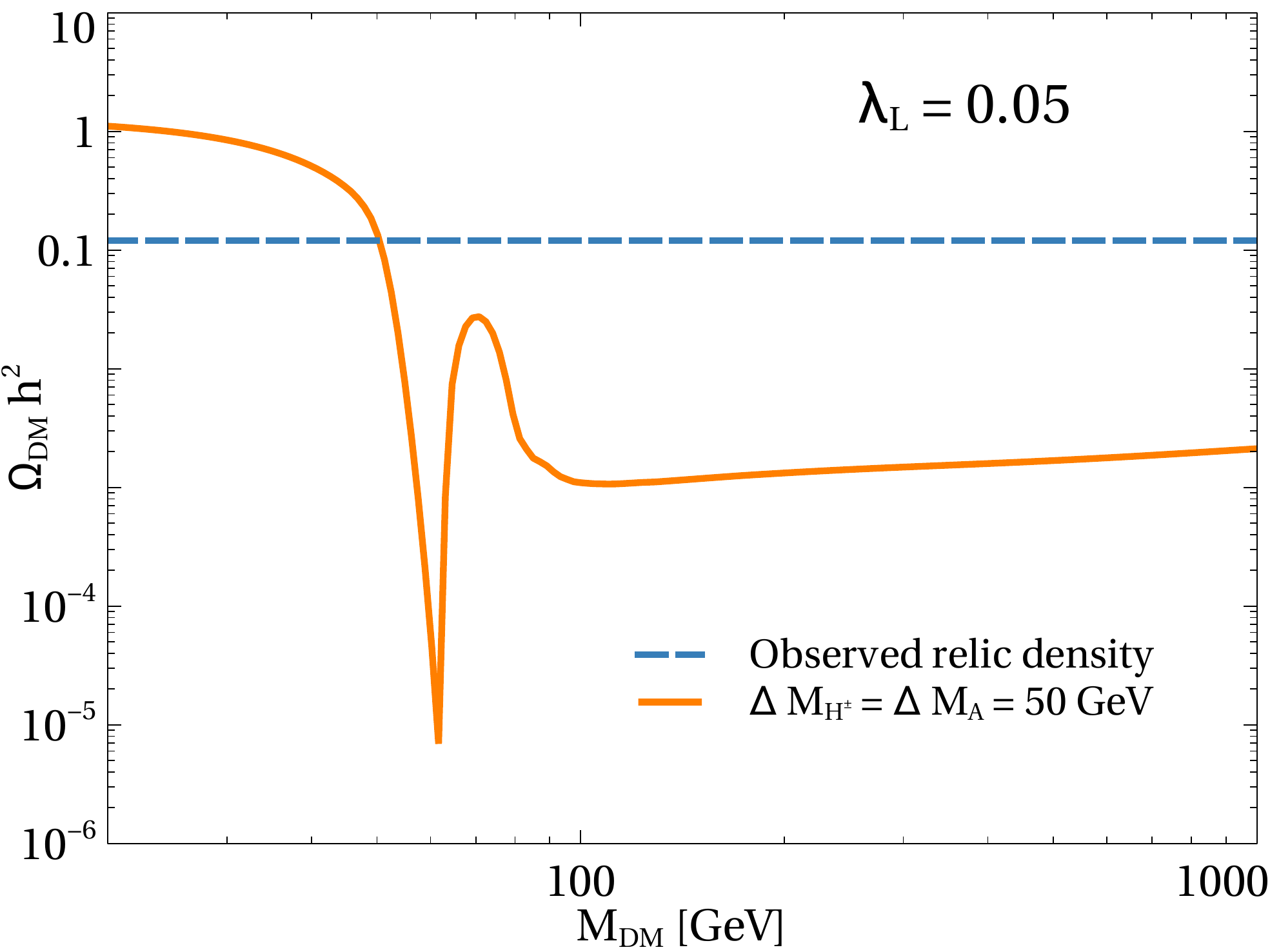} &
\includegraphics[scale=0.4]{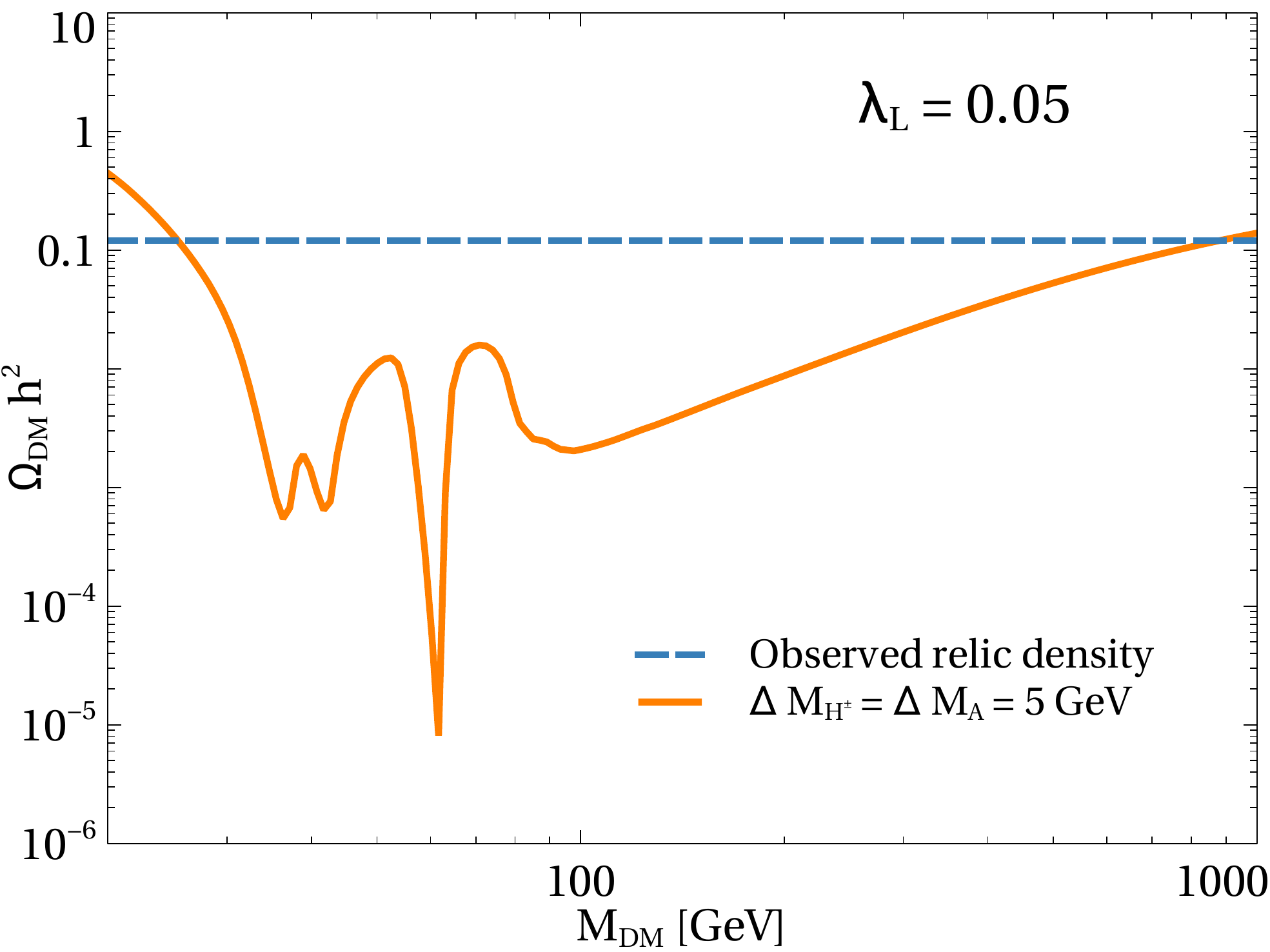} \\
\end{array}$
\caption{Relic abundance of DM in low and high mass regime for benchmark values of IHDM parameters. The values of the other parameters are: $M_h=125.5$ GeV and $\lambda_2=0.1$ respectively.}
\label{fig1}
\end{figure}
\begin{figure}[htb]
\centering
\includegraphics[scale=0.65]{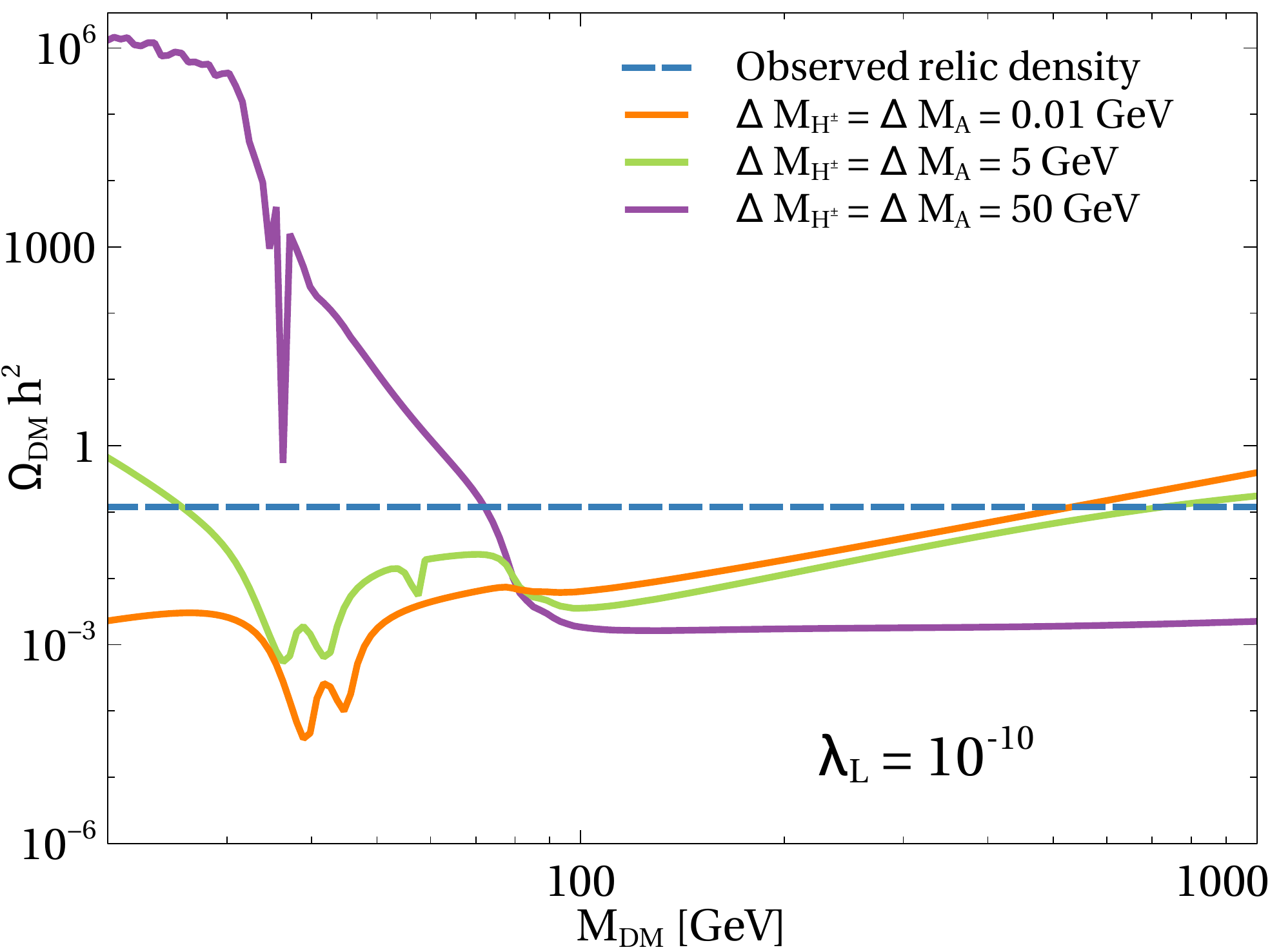}
\caption{Relic abundance of DM in low and high mass regime for vanishing Higgs portal coupling and benchmark values of mass splitting. Presence of the IDM desert is still prominent even for such low values of $\lambda_L$. The values of the other parameters are: $M_h=125.5$ GeV and $\lambda_2=0.1$ respectively.}
\label{fig2}
\end{figure}
\begin{figure}[htb]
\centering
$
\begin{array}{cc}
\includegraphics[scale=0.4]{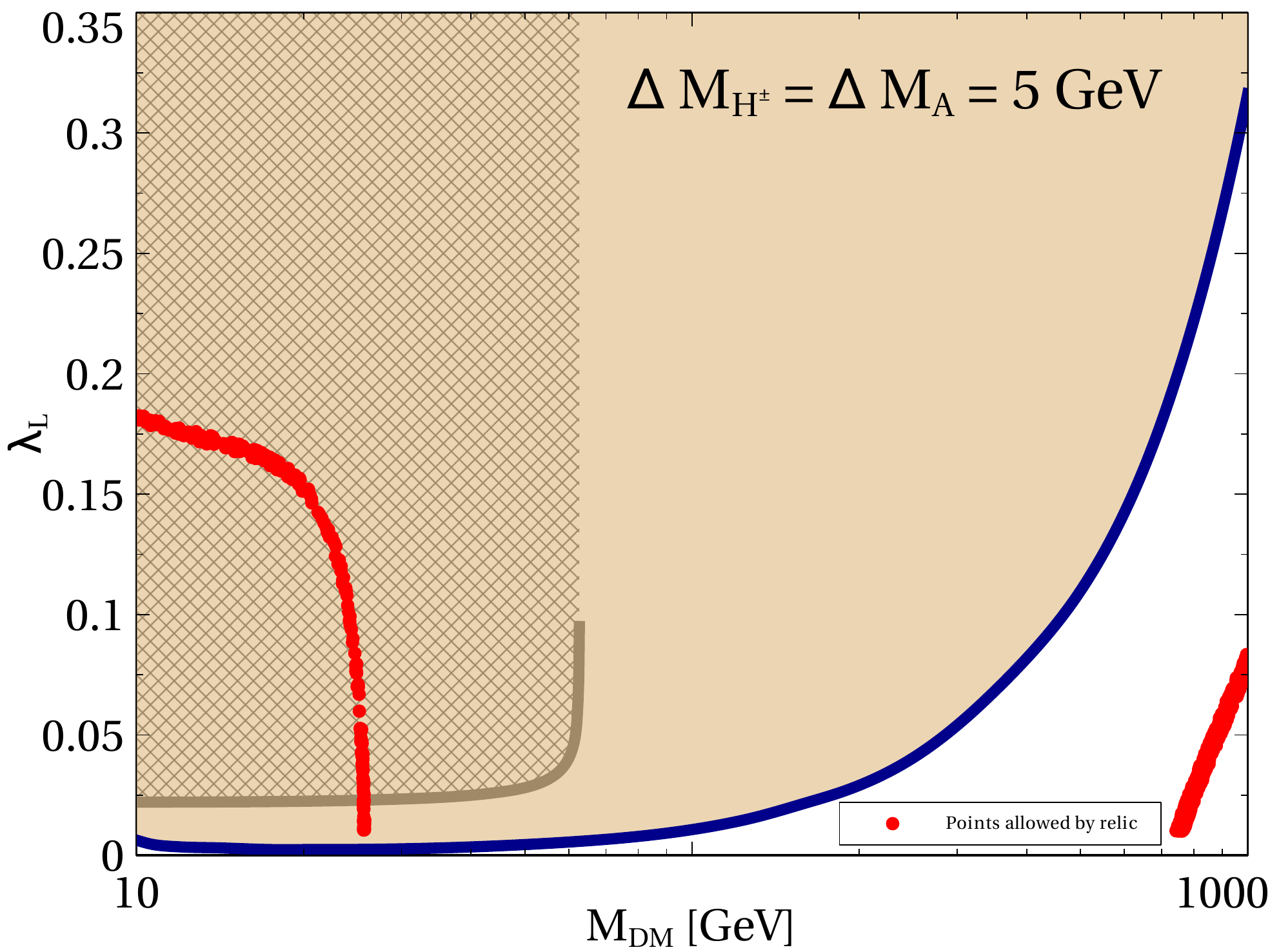} &
\includegraphics[scale=0.4]{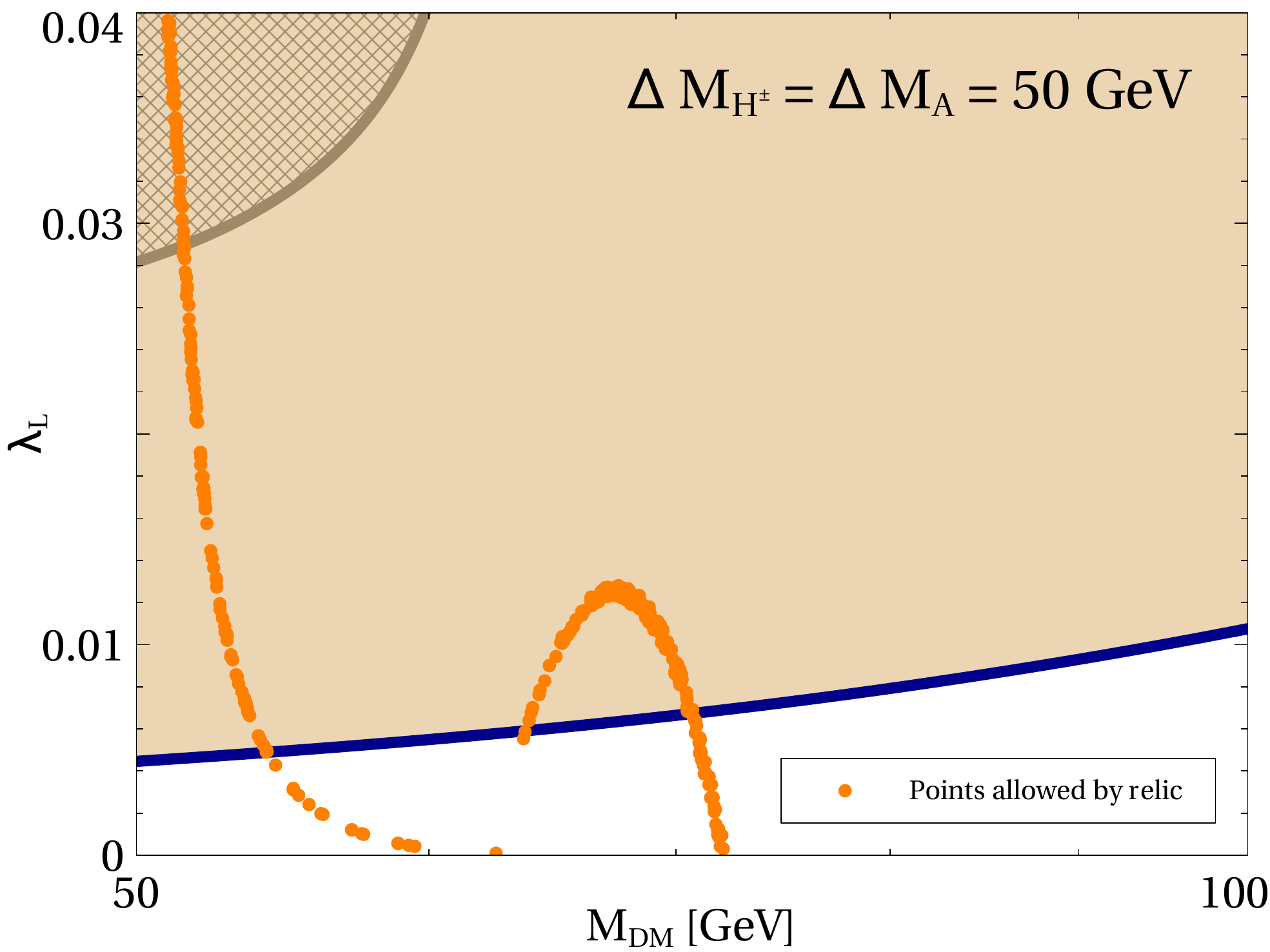}
\end{array}$
\caption{Depiction of the IDM desert in $\lambda_L$ vs $M_{\rm DM}$ plane. The range of variation for the two parameters are : $0.001 \leq \lambda_L \leq 1$ and 
	$10\,\, \rm GeV \leq M_{\rm DM} \leq 1000 \,\,\rm GeV$. The exclusion lines corresponding to XENON1T and LHC limit on Higgs invisible decay are included. The hatched region is the region disallowed by LHC limits on invisible decay of width while the uniform shaded portion is the area ruled by direct detection of XENON1T experiment.}
\label{fig3}
\end{figure}
We use the computational package micrOMEGA 4.3.1 \cite{Belanger:2013oya} to calculate the relic abundance of DM in IHDM. The relic abundance for both low and high mass regime of DM is shown in figure \ref{fig1} for specific benchmark values of IHDM parameters. After fixing the mass difference between different components of the inert Higgs doublet, the free parameters that remain are the DM-Higgs coupling $\lambda_L = (\lambda_3+\lambda_4+\lambda_5)/2$, DM mass $M_{\text{DM}}$ and the quartic self coupling of the inert Higgs $\lambda_2$. We show the variation of DM relic abundance with DM mass for two benchmark points of these IHDM parameters in figure \ref{fig1}. It can be seen that there exists a V shaped region around $M_{\text{DM}} \approx m_h/2$ which corresponds to the resonance in the s-channel annihilation of DM into the SM fermions mediated by the Higgs boson $h$. The difference in the two plots shown in figure \ref{fig1} is due to the difference in the choice of mass splitting between different components of the inert scalar doublet. In the right panel of figure \ref{fig1}, this mass difference is chosen to be smaller (5 GeV) compared to the one in the left panel (50 GeV) which increases the coannihilations between different components of the inert doublet resulting in reduced relic abundance in the low mass regime.

The mass range $M_W \leq M_{\text{DM}} \lessapprox 550 \; \text{GeV}$ typically remains disallowed from the relic abundance criteria. While the lower limit is robust, the upper limit depends upon different couplings of the model and can shift upward for different sets of parameters. For example, in the plot shown in the right panel of figure \ref{fig1}, the DM relic is satisfied for a mass around 1 TeV. If a slightly smaller value of DM-Higgs coupling $\lambda_L$ is chosen, then DM relic density can be satisfied for mass around 550 GeV also, as can be seen from earlier work on this model \cite{LopezHonorez:2006gr}. Beyond the $W$ boson mass threshold, the DM annihilations into W boson pairs become kinematically allowed, suppressing the relic abundance. The DM annihilations into $W$ boson pairs through the four point interaction diagram is given by \cite{Cirelli:2005uq}
\begin{equation}
\langle \sigma v \rangle \approx \frac{1}{64\pi M^2_{\text{DM}} g_s} \{ g^2 (3-4n^2+n^4)+16Y^4g^4_Y+8g^2 g^2_Y Y^2 (n^2-1) \}
\end{equation}
where $n=2$ is the multiplet dimension, $g_s = 2n$ and $Y$ is the hypercharge of the multiplet. Thus, the cross section decreases with increasing $M_{\text{DM}}$ and can bring the relic abundance to the required value if $M_{\text{DM}}$ is increased beyond a particular value, resulting in the above-mentioned desert $M_W \leq M_{\text{DM}} \lessapprox 550 \; \text{GeV}$ that can not satisfy the relic abundance requirement. It should be noted that DM annihilations into electroweak gauge bosons can also happen through s channel mediation of the SM Higgs and t channel mediation of heavier inert Higgs components. The difference between the behaviour of relic density versus DM mass plots shown in two panels of figure \ref{fig1} in the high mass regime is due to the difference in mass splitting. As shown in \cite{LopezHonorez:2006gr}, the annihilation into a pair of electroweak bosons receives a contribution which grows with the square of the mass difference due to which the plot in the left panel of figure \ref{fig1} shows a suppressed relic abundance in the high DM mass regime. We also check the relic abundance in the limit of vanishing Higgs portal coupling of DM $(\lambda \approx 0)$ and the results are shown in figure \ref{fig2}. The figure clearly shows the IHDM desert in between the low and high mass regime. From figure \ref{fig1} and \ref{fig2} we observe that, increasing the Higgs portal couplings and the inert Higgs mass splitting gives rise to widening of the IHDM desert. At this point, we also note that there exists a viable DM mass window between $M_W$ and 160 GeV, as pointed out by \cite{LopezHonorez:2010tb}. In spite of large annihilation cross section into electroweak gauge boson in this mass regime, generating correct relic abundance is possible due to amplitude level cancellations between different diagrams contributing to DM annihilations into electroweak gauge bosons. Since this corresponds to a very fine-tuned parameter space, we do not look into the details of it in the present work.

Instead of sticking to one possible DM-Higgs coupling, we also scan the possible values of DM-Higgs coupling $\lambda_L$ for the same choices of mass squared differences 5 GeV, 50 GeV and show the allowed region of parameter space in $\lambda_L-M_{\text{DM}}$ plane from the requirement of satisfying the correct relic abundance in figure \ref{fig3}. Apart from the relic abundance constraints on the $\lambda_L-M_{\text{DM}}$ parameter space, there also exists strict bounds from the dark matter direct detection experiments mentioned earlier. The scalar dark matter in IHDM can give rise to spin independent scattering cross section with nucleons that are being tightly constrained by the recent bounds from direct detection experiments like LUX \cite{Akerib:2016vxi}, PandaX-II \cite{Tan:2016zwf} and XENON1T \cite{Aprile:2017iyp}. For scalar dark matter considered in this work, the relevant spin independent scattering cross section mediated by SM Higgs is given as \cite{Barbieri:2006dq}
\begin{equation}
 \sigma_{\text{SI}} = \frac{\lambda^2_L f^2}{4\pi}\frac{\mu^2 m^2_n}{m^4_h M^2_{\text{DM}}}
\label{sigma_dd}
\end{equation}
where $\mu = m_n M_{\text{DM}}/(m_n+M_{\text{DM}})$ is the DM-nucleon reduced mass and $\lambda_L=(\lambda_3+\lambda_4+\lambda_5)/2$ is the quartic coupling involved in DM-Higgs interaction. A recent estimate of the Higgs-nucleon coupling $f$ gives $f = 0.32$ \cite{Giedt:2009mr} although the full range of allowed values is $f=0.26-0.63$ \cite{Mambrini:2011ik}. Similar estimates can also be found in \cite{Alarcon:2011zs, Alarcon:2012nr}. We also note that there can be sizeable DM-nucleon scattering cross section at one loop level as well, which does not depend on the Higgs portal coupling discussed above. Even in the minimal IHDM such one loop scattering can occur with charged scalar and electroweak gauge bosons in loop \cite{Klasen:2013btp}. The contributions of such one loop scattering can be kept below the latest direct detection experiments like XENON1T by choosing appropriate mass differences between the components of the inert scalar doublet \cite{Klasen:2013btp}.

One can also constrain the DM-Higgs coupling $\lambda_L$ from the latest LHC constraint on the invisible decay width of the SM Higgs boson. This constraint is applicable only for dark matter mass $m_{DM} < m_h/2$. The invisible decay width is given by
\begin{equation}
\Gamma (h \rightarrow \text{Invisible})= {\lambda^2_L v^2\over 64 \pi m_h} 
\sqrt{1-4\,m^2_{DM}/m^2_h}
\end{equation}
The latest constraint on invisible Higgs decay from the ATLAS experiment at the LHC is \cite{Aad:2015pla}
$$\text{BR} (h \rightarrow \text{Invisible}) = \frac{\Gamma (h \rightarrow \text{Invisible})}{\Gamma (h \rightarrow \text{Invisible}) + \Gamma (h \rightarrow \text{SM})} < 22 \%$$
We find it to be weaker than the latest direct detection bound from the XENON1T experiment mentioned above. This is also visible from the respective exclusion lines in figure \ref{fig3}.

Apart from direct detection experiments, DM parameter space in IHDM can also be probed at different indirect detection experiments that are looking for SM particles produced either through DM annihilations or via DM decay in the local Universe. Among these final states, photon and neutrinos, being neutral and stable can reach the indirect detection experiments without getting affected much by intermediate regions. If the DM is of WIMP type, like the one we are discussing in the present work, these photons lie in the gamma ray regime that can be measured at space based telescopes like the Fermi-LAT or ground based telescopes like MAGIC. Here we constrain the DM parameters from the indirect detection bounds arising from the global analysis of the Fermi-LAT and MAGIC observations of dSphs \cite{Ahnen:2016qkx}.

The observed differential gamma ray flux produced due to DM annihilations is given by
\begin{equation}
\frac{d\Phi}{dE} (\triangle \Omega) = \frac{1}{4\pi} \langle \sigma v \rangle \frac{J (\triangle \Omega)}{2M^2_{\text{DM}}} \frac{dN}{dE}
\end{equation}
where $\triangle \Omega$ is the solid angle corresponding to the observed region of the sky, $\langle \sigma v \rangle$ is the thermally averaged DM annihilation cross section, $dN/dE$ is the average gamma ray spectrum per annihilation process and the astrophysical $J$ factor is given by
\begin{equation}
J(\triangle \Omega) = \int_{\triangle \Omega} d\Omega' \int_{LOS} dl \rho^2(l, \Omega').
\end{equation}
In the above expression, $\rho$ is the DM density and LOS corresponds to line of sight. Thus, measuring the gamma ray flux and using the standard astrophysical inputs, one can constrain the DM annihilation into different final states like $\mu^+ \mu^-, \tau^+ \tau^-, W^+ W^-, b\bar{b}$. Since DM can not couple to photons directly, gamma rays can be produced from such charged final states. Using the bounds on DM annihilation to these final states \cite{Ahnen:2016qkx}, we show the status of IHDM for different benchmark values of parameters. 

\begin{figure}[!h]
\centering
\begin{tabular}{cc}
\epsfig{file=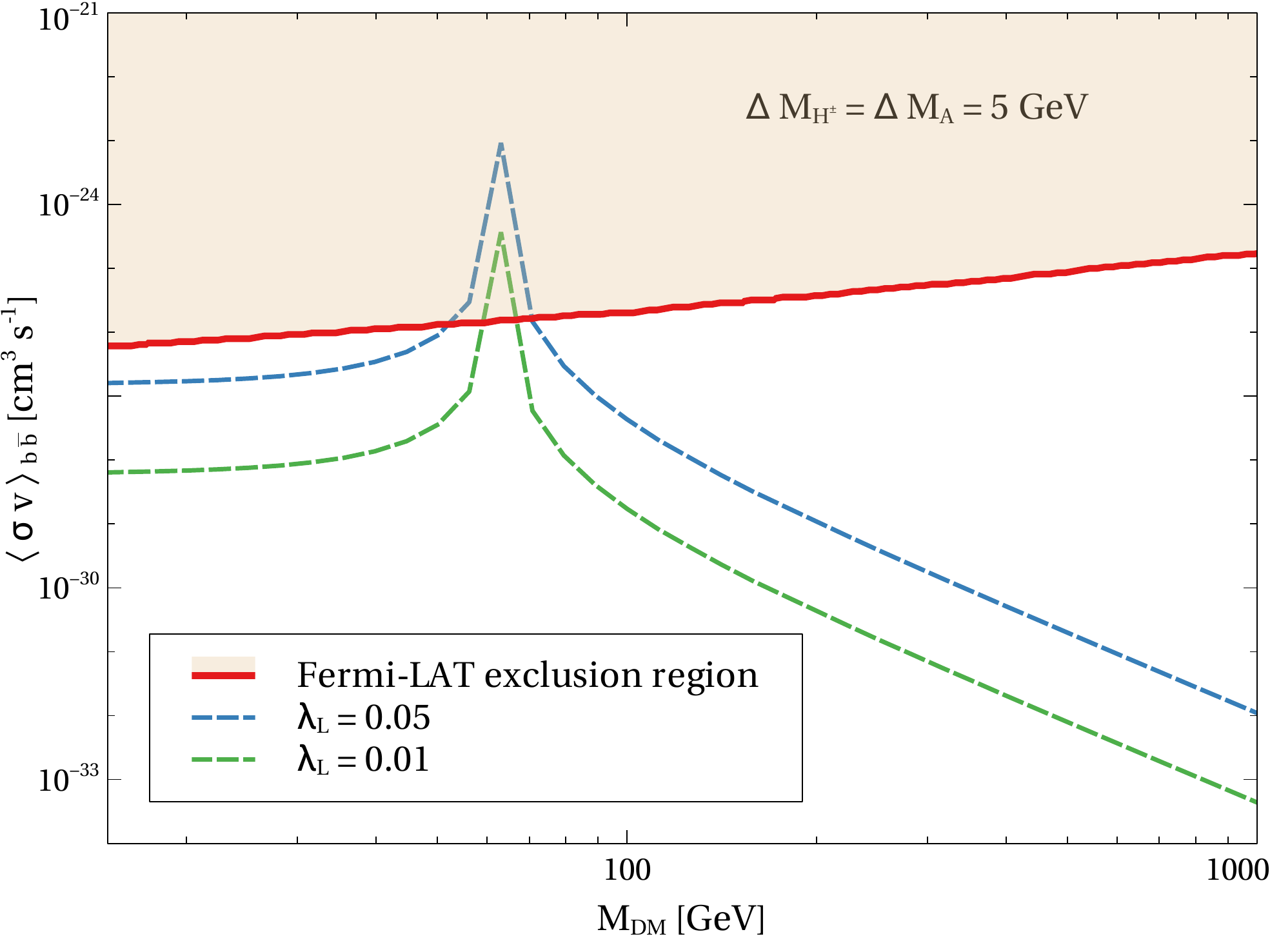,width=0.50\textwidth,clip=}
\epsfig{file=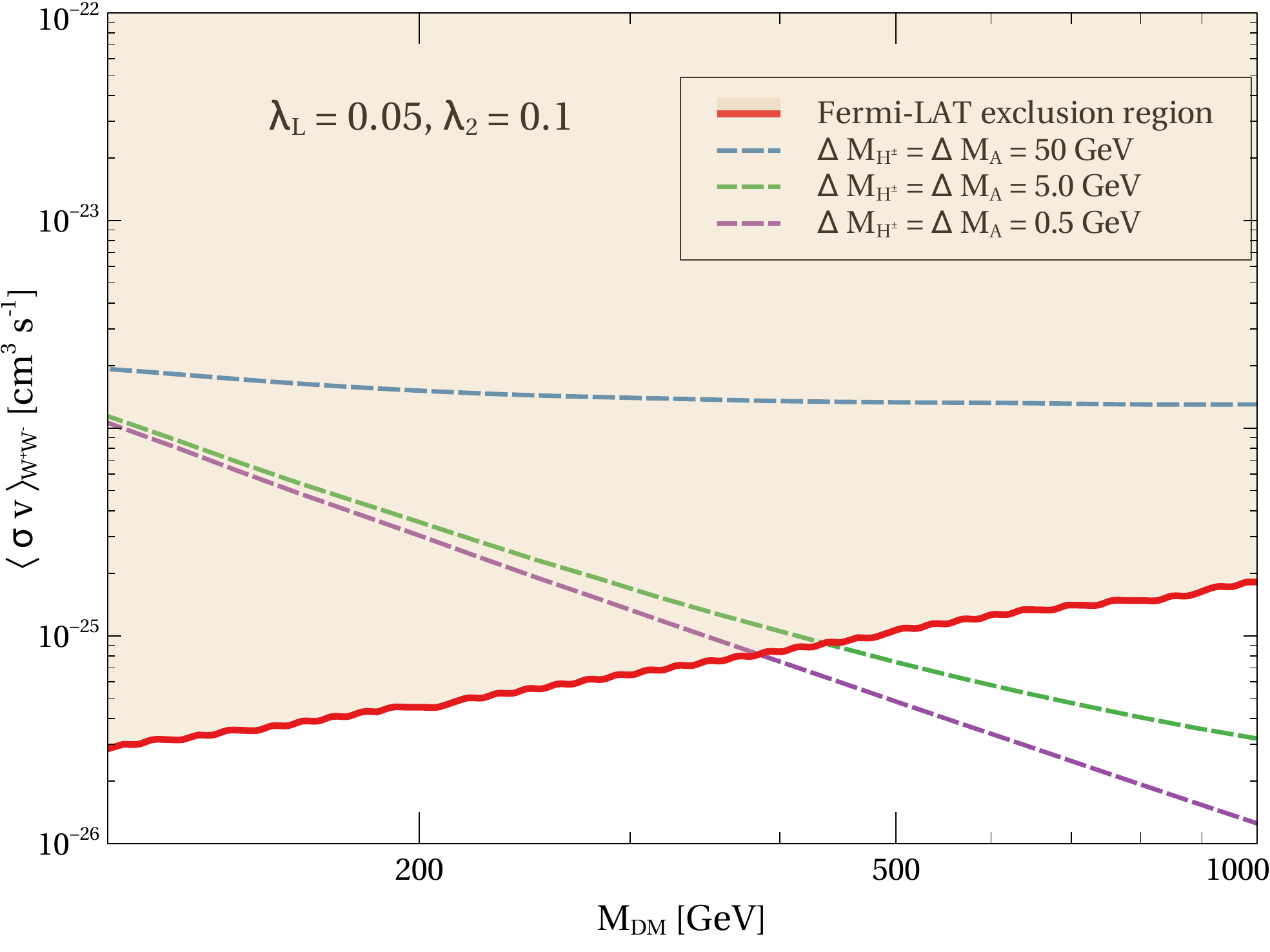,width=0.50\textwidth,clip=}
\end{tabular}
\caption{DM annihilations into $b\bar{b}$ (left), $W^+ W^-$ (right) compared against the latest indirect detection bounds of Fermi-LAT. Dark matter masses $\lessapprox 400$ GeV is ruled by annihilation to $W^+ W^-$ (right plot) for small $\Delta M_{H^\pm}$ and $\Delta M_{A}$.}
\label{fig3a}
\end{figure}

\begin{figure}[htb]
\centering
\includegraphics[scale=0.65]{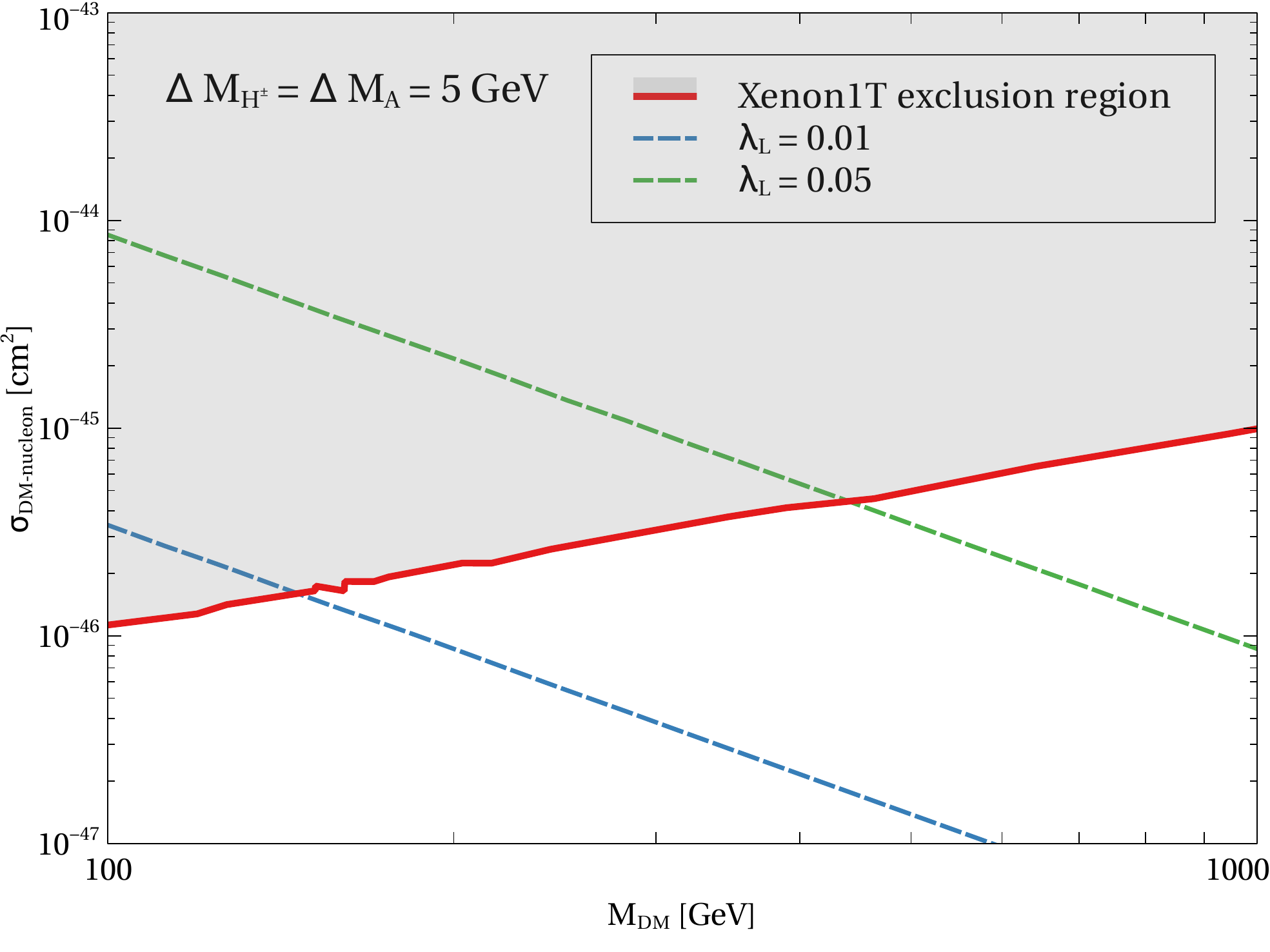}
\caption{DM-nucleon cross section and constraints from Xenon1T for different values of $\lambda_L$. Masses of DM $< 440$ GeV is ruled out for $\lambda_L = 0.05$.}
\label{fig3b}
\end{figure}

Since the heavier components of the inert doublet are not present today, therefore we do not consider the production of such SM final states from coannihilations. In the low mass regime of DM, the DM annihilations into $b \bar{b}$ pairs through SM Higgs mediation is the most relevant process that can be constrained from the indirect detection data. On the other hand, in the high mass regime, DM annihilation into $W^+ W^-$ pairs is the most relevant one. We show the DM annihilation cross section into both these final states in figure \ref{fig3a}. The left panel plot of figure \ref{fig3a} shows that the indirect detection constraints on the $b\bar{b}$ final state is weak in the high mass regime. For our cases of interest, $i.e.$ for masses of dark matter well within the IHDM desert, constraints from this channel (as well as other channels like $\mu^+ \mu^-$ and $ \tau^+ \tau^-$) are relatively weak. The annihilation cross section to $b \bar{b}$ depends very mildly on the mass splittings, but varies significantly with $\lambda_L$ as expected. On the other hand, the indirect detection bounds on the $W^+ W^-$ final state from DM annihilations heavily constrain major portions of the IHDM desert under study. For large mass splittings of $\Delta M_{H^\pm} = \Delta M_{A} = 50$ GeV, DM masses even beyond a TeV has been ruled out, as shown in the right panel of figure \ref{fig3a}. If we lower the mass splitting, even upto the order of a MeV, then also the DM mass range $M_W < m_{DM} \lessapprox 400 $ GeV is ruled out by the latest Fermi-LAT data. The annihilation cross section $\langle \sigma v \rangle_{W^+  W^-}$, does not vary much with small enough $\lambda_2$ and $\lambda_L$ but has a strong dependence on the mass splittings. Finally, it is straightforward to see that by tuning the DM-Higgs coupling, we can keep this high mass regime within XENON1T bounds, as seen from figure \ref{fig3b}. As can be seen from the right panel plot of figure \ref{fig3a}, DM masses above 400-450 GeV are allowed if the mass splitting $\Delta M_{H^\pm} = \Delta M_{A}$ is kept at $0.5-5$ GeV. Since correct relic can be produced even for purely thermal DM if the mass is approximately above 550 GeV, we particularly focus on the mass range $400 \; \text{GeV} < m_{DM} < 550 \; \text{GeV}$ which is allowed from all constraints but correct relic abundance can not be generated thermally. In the next section, we show how a non-thermal origin of DM in scotogenic extension of IHDM can generate correct relic abundance in this mass range.

After reproducing the results of IHDM in the light of the latest experimental bounds, we consider an extension of IHDM to incorporate non-zero neutrino mass which simultaneously can allow the IHDM desert $400 \; \text{GeV} \leq M_{\text{DM}} \lessapprox 550 \; \text{GeV}$ from relic abundance criteria, in the following sections.
\begin{figure}[htb]
\centering
\includegraphics[scale=0.75]{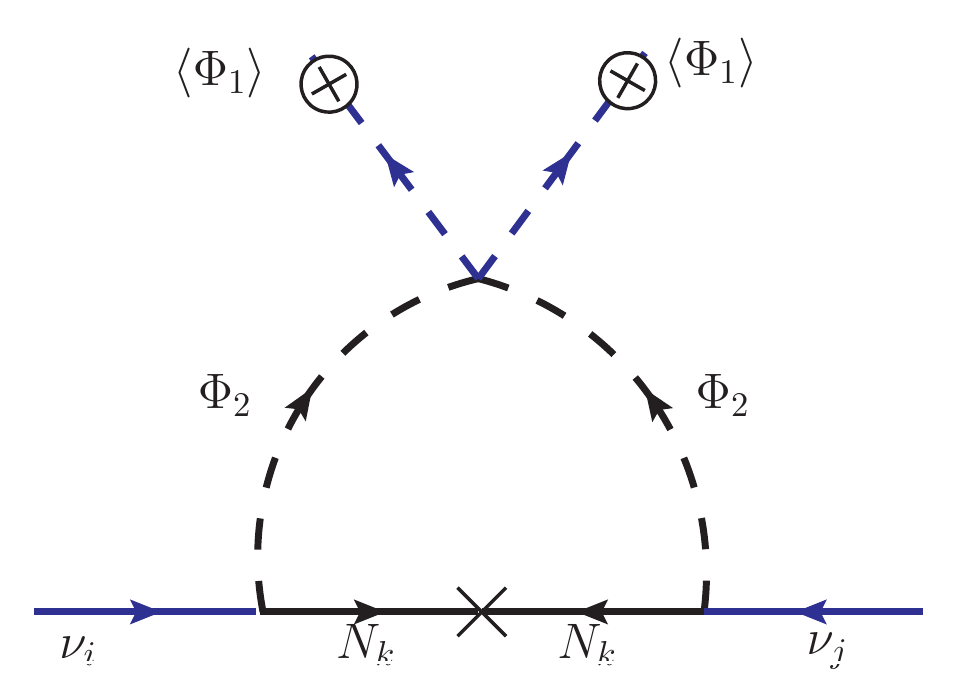}
\caption{One-loop contribution to neutrino mass}
\label{fig2a}
\end{figure}

\section{Scotogenic Extension of IHDM}
\label{sec3}
The IHDM discussed above can successfully accommodate DM but still falls short of explaining the origin of tiny neutrino masses at renormalisable level. The model can be extended by three singlet neutral fermions $N_i, i=1,2,3$ in order to generate neutrino masses. There are two different ways $N$ can generate neutrino mass. If $N \rightarrow N$ under the $Z_2$ symmetry, then they couple to the SM leptons through the usual SM Higgs $\Phi_1$ and tiny neutrino masses arise through the type I seesaw mechanism \cite{Minkowski:1977sc, Mohapatra:1979ia, Schechter:1980gr}. This does not affect the DM phenomenology of the IHDM and the origin of neutrino mass remains decoupled from the DM characteristics. A more interesting way to generate neutrino mass is through the scotogenic framework \cite{Ma:2006km} where $N \rightarrow -N$ under the $Z_2$ symmetry and hence $N$ can couple to the SM leptons only through the inert Higgs doublet $\Phi_2$. This keeps the neutrinos massless at tree level. However, at one loop level, tiny neutrino masses can arise through the diagram shown in figure \ref{fig2a}. The Lagrangian involving the newly added singlet fermion is
\begin{equation}
{\cal L} \supset  \frac{1}{2}(M_N)_{ij} N_iN_j + \left(Y_{ij} \, \bar{L}_i \tilde{\Phi}_2 N_j  + \text{h.c.} \right) \ . 
\end{equation}
The one loop neutrino mass arising from the diagram in figure \ref{fig2a} can be estimated as \cite{Ma:2006km}
\begin{equation}
(m_{\nu})_{ij} = \sum_k \frac{Y_{ik}Y_{jk} M_{k}}{16 \pi^2} \left ( \frac{m^2_R}{m^2_R-M^2_k} \text{ln} \frac{m^2_R}{M^2_k}-\frac{m^2_I}{m^2_I-M^2_k} \text{ln} \frac{m^2_I}{M^2_k} \right)
\label{numass1}
\end{equation}
Here $m^2_{R,I}=m^2_{H,A}$ are the masses of scalar and pseudo-scalar part of $\Phi^0_2$ and $M_k$ is the mass of singlet fermion $N_k$ in the internal line. The index $i, j = 1,2,3$ runs over the three fermion generations as well as three copies of $N_i$. For $m^2_{H}+m^2_{A} \approx M^2_k$, the above expression can be simply written as
\begin{equation}
(m_{\nu})_{ij} \approx \sum_k \frac{\lambda_5 v^2}{32 \pi^2}\frac{Y_{ik}Y_{jk} }{M_k} =  \sum_k\frac{m^2_A-m^2_H}{32 \pi^2}\frac{Y_{ik}Y_{jk} }{M_k}
\label{numass2}
\end{equation}
In this model for the neutrino mass to match with experimentally observed limits ($\sim 0.1$~eV), Yukawa couplings of the order $10^{-3}$ are required if the right handed
neutrino masses are as low as 1 TeV and the mass difference between $H, A$ is kept around 1 GeV. Such a small mass splitting between $H, A$ will correspond to small quartic coupling $\lambda_5 \sim 10^{-4}$. Thus, one can suitably choose the Yukawa couplings, quartic coupling $\lambda_5$ and the right handed neutrino masses in order to arrive at sub eV light neutrino masses.
\section{Dark Matter in Scotogenic IHDM}
\label{sec4}
In the scotogenic version of IHDM, there can be two different DM candidates: either the lightest $N_i$ or the lightest neutral component of $\Phi_2$, depending on whichever is lighter. We consider the latter possibility and study the changes in DM phenomenology of IHDM after the introduction of the singlet fermions. This requires all the singlet neutral fermions $N_i$ to be heavier than the scalar DM candidate. Now, there can be two different ways the scalar DM phenomenology can be significantly different from that of IHDM. We discuss them below.
\subsection{Purely Thermal DM}
The calculation of DM relic abundance assuming only thermal contributions is similar to IHDM except the fact that there exists another annihilation channel due to the presence of singlet neutral fermion $N$. The DM can self annihilate into the light neutrinos through t-channel exchange of $N$. The contribution of this diagram however, remains suppressed due to p-wave suppression and also due to the requirement of satisfying tiny neutrino mass data, in comparison to other annihilation channels. As can be seen from the simplified formula for neutrino masses given in \eqref{numass2}, the Yukawa couplings can be of order one $Y \sim \mathcal{O}(1)$ only if the mass difference $m_A-m_H$ is taken as low as $\mathcal{O}(100 \; \text{keV})$, typical kinetic energy of DM particles. This lower limit on the mass difference is required to avoid tree level inelastic scattering of DM off nuclei mediated by $Z$ boson \cite{Arina:2009um}. However, such small mass difference between $H, A$ makes the gauge boson mediated coannihilations so efficient that there is hardly any strong dependence of DM relic abundance on the new t-channel annihilation channel mediated by heavy neutrinos \cite{Borah:2017dqx}. Also, even if this new annihilation channel has a sizeable contribution to DM abundance compared to other channels, like in generic lepton portal DM models \cite{Chang:2014tea, Bai:2014osa}, it does not help in generating new allowed DM masses in the IHDM desert $400 \,\rm GeV \leq M_{\text{DM}} \lessapprox 550 \; \text{GeV}$ mentioned above. Therefore, we do not perform such a calculation in details and move on to the non-thermal DM scenario which can generate correct relic abundance even if the DM mass falls in this IHDM desert.


\subsection{Mixture of Thermal and Non-thermal DM}
In this section we discuss the possibility of reviving the IHDM desert by bringing the relic abundance of DM in the mass range $400 \; \text{GeV} \leq M_{\text{DM}} \lessapprox 550 \; \text{GeV}$ to the observed range through a non-thermal contribution to the relic abundance. This is possible by virtue of a late time decay of a heavy particle into DM which can bring the under-abundant DM in the IHDM desert to match with the observed limit from the Planck mission. Since the DM with mass in the IHDM desert undergoes thermal freeze-out, the decay of the heavy particle should occur after the DM freeze-out to have an impact on its abundance. Since thermal freeze-out of WIMP dark matter occurs typically at temperatures around a GeV, one expects the lifetime of the decaying particle to be more than a microsecond. Although there are two more particles in IHDM namely $H^\pm, A$ which heavier than DM, their lifetime will be much shorter than a microsecond for all realistic mass differences between the inert Higgs doublet components. This makes the lightest neutral singlet fermion $N$ in the scotogenic extension of IHDM, a more natural choice for a long-lived particle decaying into DM after the thermal freeze-out of DM.

Here we consider the production of $N$ to be negligible and hence its decay is out of equilibrium. On the other hand, the DM particle we consider here can have both thermal and non-thermal production. Summary of such a hybrid setup of both thermal and non-thermal production of DM can be found in \cite{Gelmini:2010zh}. The late decay of a heavy field, for example, a moduli field in supersymmetric scenarios \cite{Moroi:1999zb}, can reheat the Universe to a low reheating temperature. In such low temperature reheating scenarios, DM can get a production enhancement from the decay of such heavy fields and at the same time, the release of entropy from such decay can also suppress the DM abundance. Therefore, in such non-standard cosmology scenarios, the final relic density of DM can be larger or even smaller compared to the standard cosmology case. For example, in the standard scenario, the dark matter can be overproduced due to small annihilation channels. In such a case, if late decay of a heavy particle produces lots of radiation after DM freeze-out, it injects entropy into the system and decreases the abundance of DM through the mechanism of entropy dilution \cite{Scherrer:1984fd}. Here we consider a scenario where DM is thermally under-produced and a non-thermal origin due to the decay of a heavier particle can bring its abundance into the observed range. Within supersymmetric models, non-thermal production of typical WIMP candidates like Wino-Higgsino from late decay of a scalar field without giving rise to the gravitino overproduction was discussed in \cite{Endo:2006ix}.

In our work, following the approach of \cite{Drees:2006vh}, the number densities of DM and the lightest singlet fermion $N$ can be calculated by solving the coupled Boltzmann equations given by
\begin{equation}
\frac{dn_{\rm DM}}{dt}+3Hn_{\rm DM} = -\langle \sigma v \rangle (n^2_{\rm DM} -(n^{\rm eq}_{\rm DM})^2) +N_{\rm DM} \Gamma_N n_N
\end{equation}
\begin{equation}
\frac{dn_{N}}{dt}+3Hn_{N} = -\Gamma_N n_N
\end{equation}
Here, $N_{\rm DM}$ is the average number of particles produced from a single decay of $N$ and $\Gamma_N$ is the decay width of $N$. Assuming that these singlet neutrinos do not contribute dominantly to the total energy budget, we can take the comoving entropy density ($g_{\star s}$) and the comoving energy density ($g_\star$) to be approximately constant. Further, we assume that almost all of $N$ decays during the radiation dominated epoch. With these assumptions, one can analytically solve the Boltzmann equation for $n_N$ above. Writing the above equations in terms of $Y_{\rm DM}=\frac{n_{\rm DM}}{s}, Y_{N}=\frac{n_N}{s}$ with $s = \frac{2\pi^2}{45}g_{*s} T^3$ being the entropy density and changing the variable from time $t$ to $x=M_{\text{DM}}/T$, we get
\begin{equation}
\dfrac{dY_{\rm DM}}{dx}=-\dfrac{\langle\sigma v\rangle\,s}{H\,x}\left(Y_{\rm DM}^2-(Y^{\rm eq}_{\rm DM})^2\right)+\frac{N_{\rm DM} \Gamma_N}{H \,x} Y_N
\end{equation}
\begin{equation}
\dfrac{dY_N}{dx}= -\frac{\Gamma_N}{H\,x} Y_N
\end{equation}
The equation for $Y_N$ can be solved analytically to give 
\begin{equation}
Y_N(x) = Y_N (x_F) \text{exp} \left(-\dfrac{r}{2}\left(x^2-x_F^2\right)\right)
\end{equation}
Here, $x_F=M_{\rm DM} / T_F$ is the point of freeze-out and usually takes a value of $\mathcal{O}(20)$. Also, $r=\dfrac{\Gamma_N}{H\,x^2}=\dfrac{\Gamma_N\,M_{\rm Pl}}{\pi M_{\rm DM}^2}\sqrt{90/g_{\star}}$, depends on the decay width of the mother particle. $Y_N(x_0)$ however, depends on the initial abundance of $N$ and is treated as a free parameter. Using this solution for $Y_N$, the equation for $Y_{\rm DM}$ can be rewritten as
\begin{equation}
\dfrac{dY_{\rm DM}}{dx}=-\dfrac{\langle\sigma v\rangle\,s}{H\,x}\left(Y_{\rm DM}^2-(Y^{\rm eq}_{\rm DM})^2\right)+N_{\rm DM}\,r\,x\,Y_N(x_0)\text{exp}\left(-\dfrac{r}{2}\left(x^2-x_0^2\right)\right)
\label{y_eq}
\end{equation}
In the above expressions, $g_*, g_{*s}$ are the relativistic degrees of freedom that contribute to the total radiation density and entropy density of the Universe respectively. Solving the above equation \eqref{y_eq} numerically will give the abundance of $\rm DM$ in the present Universe.

The most dominant decay mode to which the singlet neutrino can decay to is $N \rightarrow \nu \,\,\rm H$, and this is the main production channel of our dark matter candidate. Furthermore, we should be careful about two more things. The decay of the singlet neutrino should not occur during or after the epoch of the Big Bang Nucleosynthesis (BBN), because this decay can release entropy thereby disrupting the proportion of abundances of the light elements which are very precisely measured and found to be consistent with the standard $\Lambda$CDM cosmology. So, to be on safe side, the decay lifetime of the sterile neutrino ($\tau_N$) should be less than around 1s. This gives an lower bound on the decay width of $N$, i.e. $\Gamma_N \geq \Gamma_{N,\rm min} \equiv 6.58\times 10^{-25}$ GeV. Also, the decay of the sterile neutrino should contribute mostly after the usual thermal freeze-out of our dark matter candidate, since, otherwise, the decay products will get diluted within the thermal plasma failing to give the necessary contribution towards its relic density. This consideration gives an upper bound on $\Gamma_N$. This can be written in terms of the freeze-out parameter $x_0$ as $\Gamma_N \leq \Gamma_{N,\rm max}\equiv \dfrac{M_{\rm DM}^2}{x_0^2}\times 10^{-18}$ GeV.

It should be noted that the above analysis is done with the assumption that entropy per comoving volume remains constant. Although the decay of $N$ can release entropy, we can neglect it in the assumption that singlet neutrino contribution to the total energy density was small in the early Universe, as stated above. This is similar to the framework presented by the authors of \cite{Drees:2006vh}. In the most general case, another Boltzmann equation arises, corresponding to the radiation content of the Universe which receives contribution from the decay of the mother particle. Such a general analysis can be found in \cite{Giudice:2000ex}. On the other hand, such entropy release due to the late decay of heavy particles can affect the freeze-out abundance of other particles, as pointed out by \cite{Scherrer:1984fd}. Since in our case, the mother particle decays before BBN temperature, therefore it does not affect the abundance of light nuclei. However, in SuperWIMP dark matter scenarios \cite{Feng:2003uy}, such effects are non-trivial as the life-time of the mother particle is much larger than the BBN time scale.

\begin{figure}[h!]
\centering
\includegraphics[scale=0.52]{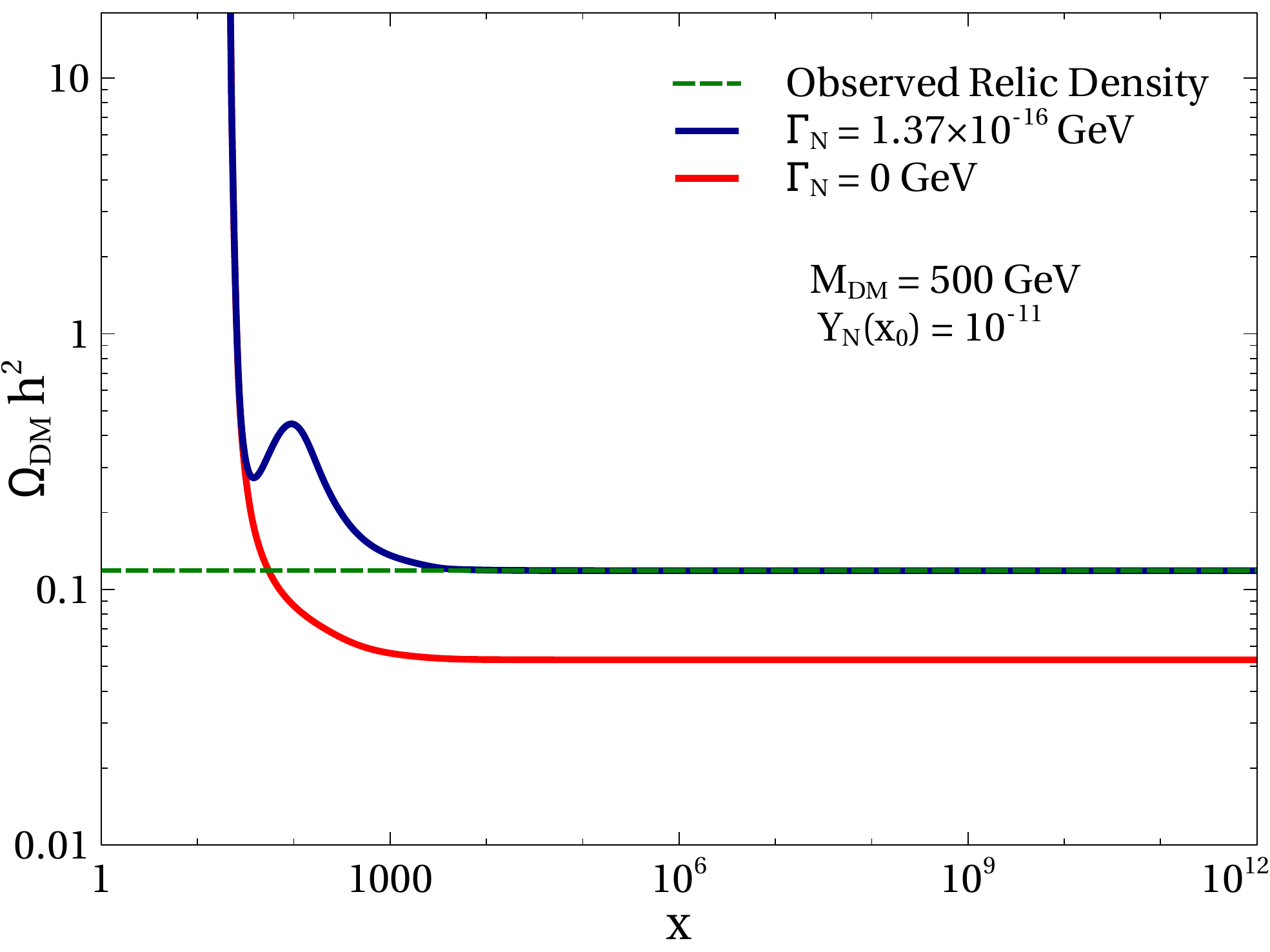}
\caption{Allowed benchmark point with mass of dark matter taken to be 500\,GeV. Corresponding values of $\Gamma_N$ required to satisfy the relic density is also given. Other parameters that contribute towards the relic density are kept fixed : $\Delta M_{{H}^\pm} = \Delta M_A = 5$ GeV, $\lambda_L=0.05,\,\,\lambda_2=0.1$, $M_{h}=125.5$ GeV. The red line shows the under-abundant (thermal only) scenario with $\Gamma_N=0$ GeV.}
\label{relic-fig}
\end{figure}

We solve the differential equation mentioned in equation \eqref{y_eq} and find the comoving number density (and consequently the relic density, $\Omega_{\rm DM}\,h^2$) as a function of the parameter $x$. In keeping with the direct and indirect detection constraints as depicted in figure \ref{fig3a}, a suitable benchmark point with $M_{DM} =$ 500 GeV is chosen. As seen from figure \ref{fig1} and figure \ref{fig2}, for this mass the dark matter is under-abundant in the pure IHDM irrespective of the choice of parameters. Although the two plots shown in figure \ref{fig1} and \ref{fig2} correspond to fixed couplings $\lambda_L, \lambda_2$, it is well known from IHDM studies that for this benchmark mass of dark matter, correct relic abundance can not be generated. Over and above this under-abundant thermal abundance, if the dark matter can be further produced from the decay of another particle, as we propose in this work, we hope the dark matter to satisfy the observed relic density. This is exactly what we observe in figure \ref{relic-fig} where we show that for this benchmark DM mass in the IHDM desert, we can have correct relic abundance if the initial abundance $(Y_N (x_0))$ and the decay width $(\Gamma_N)$ of the decaying mother particle are appropriately chosen. It should be noted that in the plots shown in figure \ref{relic-fig} and subsequent ones, the label $\Omega_{\rm DM}\,h^2$ in the y-axis denotes the relic density of the DM particle, obtained from solving the Boltzmann equation \eqref{y_eq}. It is generated by numerically solving this Boltzmann equation \eqref{y_eq} for benchmark values of input parameters. To compare the differences with pure IHDM, the corresponding $\Omega_{\rm DM}\,h^2$ versus $x$ plot is also shown which is denoted by $\Gamma_N = 0$ meaning the absence of any decay of mother particle into DM. It can be easily seen from the plots shown in figure \ref{relic-fig} that, the pure IHDM cases have under-abundant DM while the one with scotogenic extension can satisfy the correct relic abundance.

\begin{figure}[h!]
	\centering
	\includegraphics[scale=0.55]{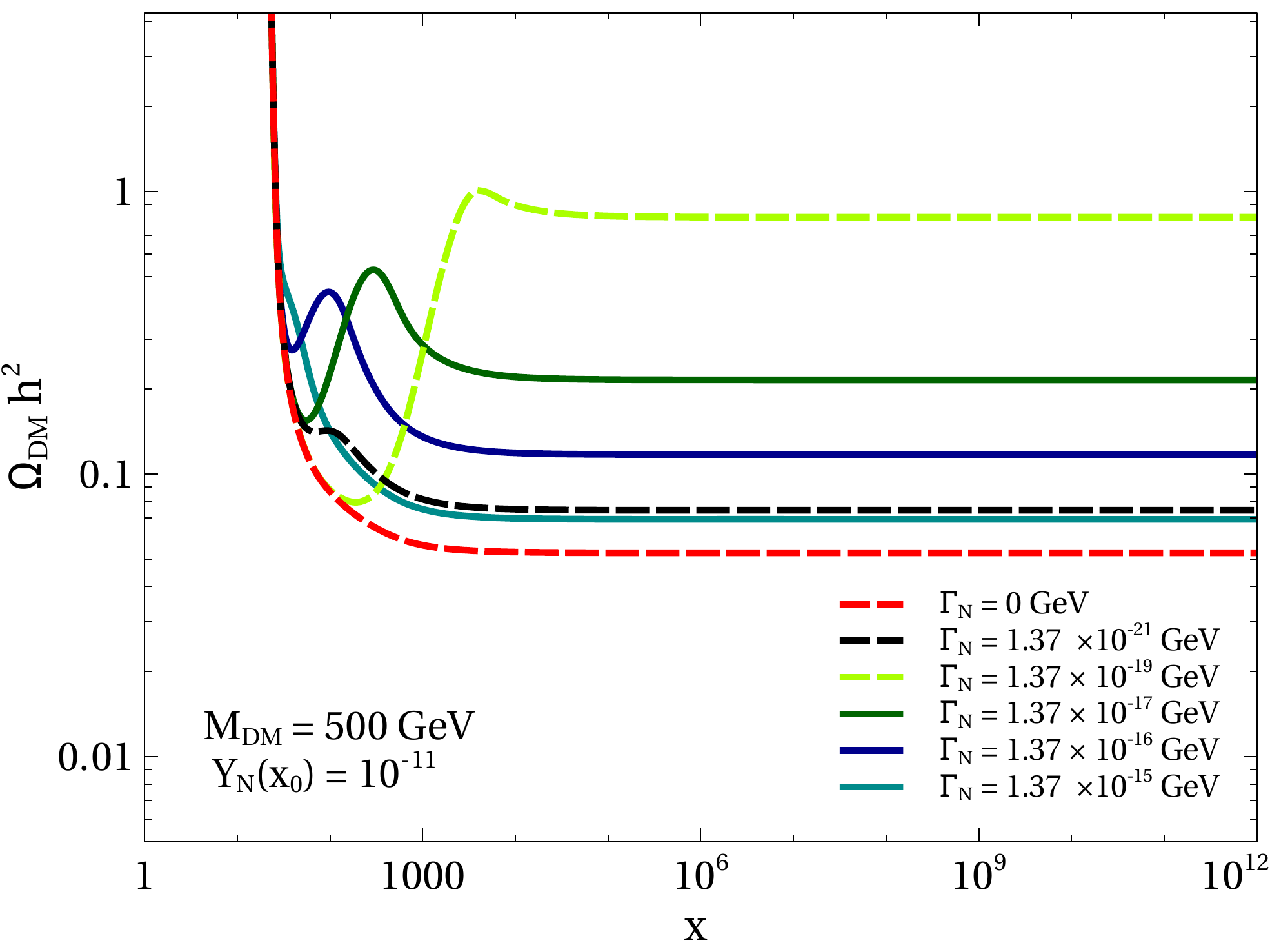}
	\caption{Variation of relic density with different values of the parameter $\Gamma_N$. Other parameters that contribute towards the relic density are kept fixed : $\Delta M_{{H}^\pm} = \Delta M_A = 5$ GeV, $\lambda_L=0.05,\,\,\lambda_2=0.1$, $M_{h}=125.5$ GeV.}
	\label{r_vary}
\end{figure}

Now, let us try to understand how the relic density changes as a function of the other control parameters $\Gamma_N$ and $Y_N(x_0)$. For that we first plot the variation of $\Omega_{\rm DM}h^2$ with different values or $\Gamma_N$. It is shown in figure \ref{r_vary}. With $\Gamma_N=0$ GeV, we find our older solution with no increase in the relic density of the dark matter which corresponds to the pure IHDM case. As we increase $\Gamma_N$ from here we see, that the final abundance (i.e. where $\Omega_{\rm DM}h^2$ flattens out to a constant value) increases first, but then with a further increment of $\Gamma_N$, it begins to decrease. This behaviour is expected because in the Boltzmann equation \eqref{y_eq}, the second term on the right hand side (RHS) contains two different types of contribution from the decaying singlet neutrino. One contribution is linear in $\Gamma_N$ whereas the other comes within the exponential function. So, if we confine $\Gamma_N$ to a range of small values, we should expect an increase in the final number relic density with and increase in $\Gamma_N$. But if we keep on increasing $\Gamma_N$, then at some point, the exponentially decaying term will start dominating over the linearly increasing term. This will result in decreasing values of the final abundance with subsequent increase of $\Gamma_N$. 

\begin{figure}[h!]
	\centering
	\includegraphics[scale=0.5]{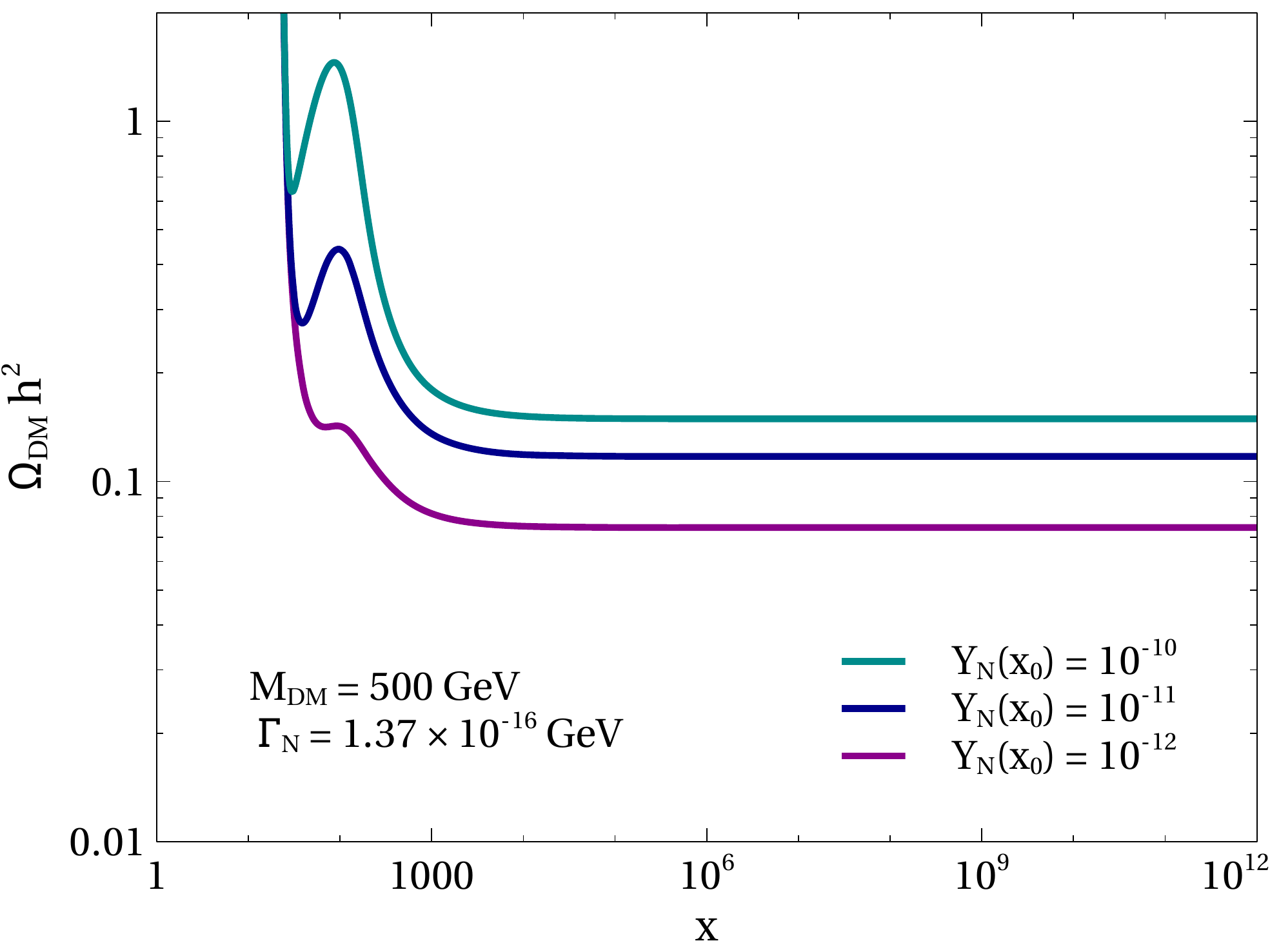}
	\caption{Variation of relic density with different values of the parameter $Y_N(x_0)$. Other parameters that contribute towards the relic density are kept fixed : $\Delta M_{{H}^\pm} = \Delta M_A = 5$ GeV, $\lambda_L=0.05,\,\,\lambda_2=0.1$, $M_{h}=125.5$ GeV.}
	\label{y_vary}
\end{figure}

The variation of the final relic density with changing $Y_N(x_0)$ is straightforward and is plotted in Fig. \ref{y_vary}.
With the increasing initial abundance of the mother particle, the final abundance of dark matter increases as expected.

\begin{figure}[h!]
	\centering
	\includegraphics[scale=0.5]{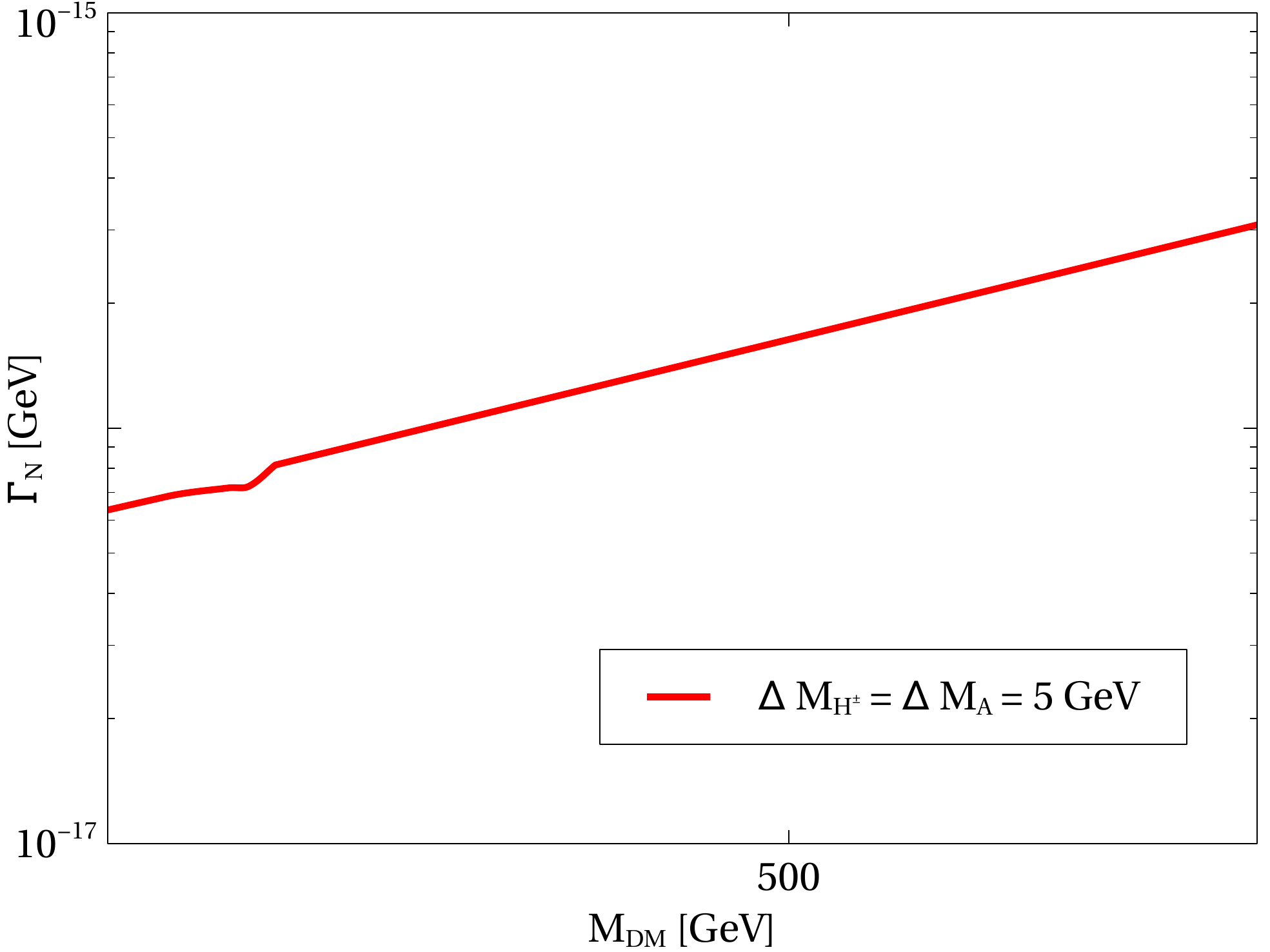}
	\caption{Points satisfying relic density ($0.1166 \leq \Omega_{\rm DM}h^2 \leq 0.1206$) plotted in M$_{\rm DM}$ vs $\Gamma_{N}$ plane. $\Gamma_N$ was varied between $\Gamma_{N,\rm min}$ and $\Gamma_{N,\rm max}$ derived earlier. $M_{\rm DM}$ was varied between 440 GeV and 1000 GeV.}
	\label{scan}
\end{figure}

To get a more complete picture we also performed a scan for all possible combinations of M$_{\rm DM}$ and $\Gamma_N$ that can give the correct relic (and are also allowed by direct and indirect detection). $\Gamma_N$ was bounded by the upper and lower limits that we derived earlier during the scan. The scans were performed keeping $\lambda_L=0.05,\,\,\lambda_2=0.1$ and the allowed mass splitting namely $\Delta M_{H^\pm}=\Delta M_{A}$ = 5 GeV. In keeping with the direct and indirect constraints $M_{DM} > 440$ GeV was chosen. The points in $\Gamma_N-M_{\rm DM}$ plane satisfying these different constraints are shown in figure \ref{scan}.

\begin{figure}[h!]
	\centering
	\includegraphics[scale=0.52]{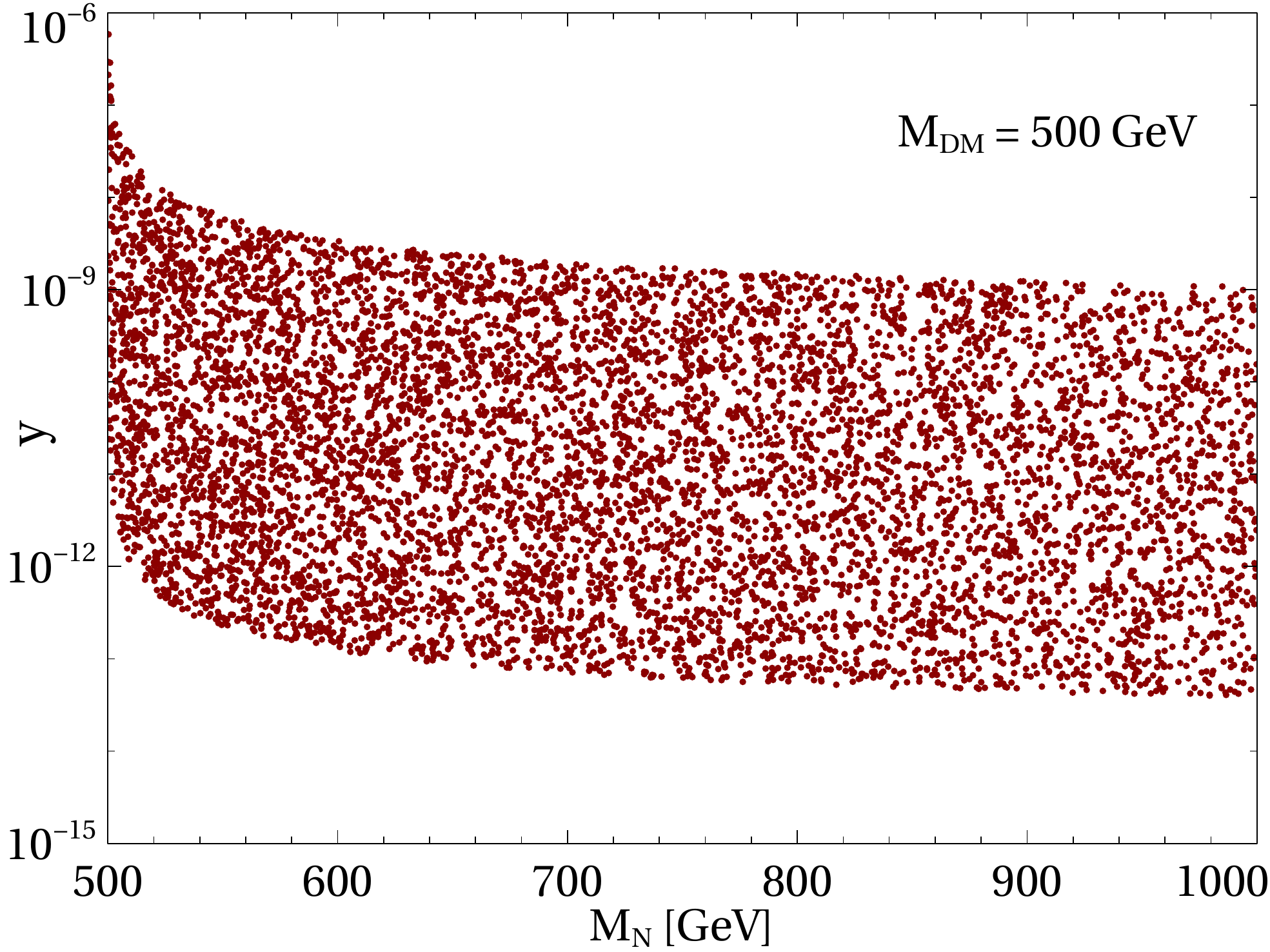}
	\caption{Allowed region in the $y-M_{\rm N}$ plane arising from the constraints on $\Gamma_N$, shown for the benchmark dark matter mass.}
	\label{yuk_mn}
\end{figure}

After studying the possibility of reviving the IHDM desert from correct dark matter relic abundance criteria in the scotogenic framework, we also check the possible implications of this scenario in the neutrino sector. Since light neutrino mass depends upon the masses of singlet neutral fermions and their couplings with DM, as discussed earlier, we can correlate neutrino mass with the decay width of the lightest singlet fermion responsible for generating correct DM relic abundance in the IHDM desert. The limit on $\Gamma_N$ we discussed above can be translated to get some idea about the allowed range of the lightest singlet fermion mass and the Yukawa coupling governing the decay of the same to a neutrino and dark matter. The corresponding decay width of $N$ is given by
\begin{equation}
\label{decay_N}
\Gamma_N = \dfrac{y^2}{4\pi}M_{\rm N}\left(1-\left(\dfrac{M_{\rm DM}}{M_{\rm N}}\right)^2\right)
\end{equation}
Using this in the expression of $\Gamma_N$ and imposing the limits on it derived from the 
BBN constraints as well from the requirement that decays should be most dominant after the thermal freeze-out of DM, we get an allowed region in the $y-M_{\rm N}$ plane. The dark matter mass has been fixed to the benchmark value of 500 GeV. The result is shown in figure \ref{yuk_mn}. It can be seen from the plots shown in this figure that for singlet fermion masses upto 1 TeV, the requirement of a long lived singlet fermion constrains the Yukawa coupling to be very small $y \leq 10^{-9}$. As can be seen from the one-loop neutrino mass formula given in \eqref{numass1}, such tiny couplings will practically have a negligible contribution to light neutrino masses. This will in fact, make the lightest right handed neutrino decoupled from the neutrino mass generation mechanism effectively giving rise to a scotogenic model with only two right handed neutrinos. This will result in one almost massless and two massive light neutrinos which can be tested at experiments like neutrinoless double beta decay which are sensitive to the absolute mass scale of neutrinos.

\section{Results and Conclusion}
\label{sec5}
We have studied the possibility of generating correct relic abundance of dark matter in the inert Higgs doublet model for the range of dark matter mass which was previously known to be inconsistent with the requirement of matching the observed relic abundance. The abundance of DM due to the usual freeze-out mechanism remains under-abundant in the IHDM desert which roughly corresponds to the mass range $400\, \rm GeV \leq M_{\text{DM}} \lessapprox 550 \; \text{GeV}$. We then check the constraints from direct and indirect detection experiments and find that even if relic abundance in this desert is satisfied due to a non-thermal source, the latest indirect detection constraints from Fermi-LAT can rule out a large portion of it $M_W \leq M_{\text{DM}} \lessapprox 400 \; \text{GeV}$ leaving the allowed region $ M_{\text{DM}} \in (400-550) \; \text{GeV}$, which we intend to revive by producing the correct relic abundance. For this purpose we have studied a minimally extended version of IHDM which not only revives the IHDM desert to generate correct DM abundance but also generates light neutrino masses, which remains unaddressed in the original IHDM. We incorporated three singlet neutral fermions into IHDM which are also odd under the in-built $Z_2$ symmetry of the model. This allows the possibility of having one-loop neutrino mass in the scotogenic fashion with the dark matter particle going inside the loop. We assume the lightest of these singlet neutral fermions $N$ to be long lived such that DM can have a non-thermal production due to the late decay of $N$.  We show that for suitable values of decay widths and initial abundance of $N$, one can successfully generate correct DM abundance in the IHDM desert. Since the low mass regime of IHDM is getting very much constrained due to null results at direct detection, the possibility of reopening this forbidden mass window within this minimal setup should be able to initiate further exploration of this minimal DM model from different experimental point of view, starting from indirect detection to collider searches. We also check the implications of such a scenario from the constraints we impose on the decay width or lifetime of $N$: that it should decay after the thermal freeze-out of DM and before BBN. This restricts the mass and Yukawa couplings of $N$ with DM and neutrinos and give rise to a negligible mass of the lightest neutrino. This will undergo scrutiny at experiments which can probe the absolute mass scale of neutrinos for example, neutrinoless double beta decay.

We have confined ourselves to a simple and minimal setup in this work to show the new viable region of IHDM parameter space, previously discarded due to under-abundant DM from thermal freeze-out mechanism. This work can be suitably extended to a richer framework where the initial abundance of $N$, its tiny couplings with DM can be explained naturally. It will also be interesting to study the collider and indirect detection prospects of this newly available DM mass window. We leave such a detailed study to future works.

\acknowledgments
AG would like to acknowledge the hospitality provided by the department of physics, IIT Guwahati during his visit in January 2017 when this work was initiated. AG also acknowledges Department of Atomic Energy (DAE), Govt. of India for their financial assistance. AG is grateful to Dr. Mehedi Masud for his discussions regarding some aspects of numerical calculations used in this work. DB acknowledges the support from IIT Guwahati start-up grant (reference number: xPHYSUGIITG01152xxDB001) and Associateship Programme of IUCAA, Pune.


\end{document}